%% file: ms.tex
\documentclass[ejs]{imsart}

\usepackage{amsfonts}
\usepackage{multirow}
\RequirePackage{amsthm,amsmath}
\RequirePackage[numbers]{natbib}
\RequirePackage[colorlinks,citecolor=blue,urlcolor=blue]{hyperref}
\RequirePackage{graphicx}

\arxiv{2207.14753}

\startlocaldefs
\numberwithin{equation}{section}
\theoremstyle{plain}
\newtheorem{thm}{Theorem}[section]

\endlocaldefs

\newcommand{\argmin}[1]{{\underset{#1}{\textnormal{argmin}} \, }}
\newcommand{\argmax}[1]{{\underset{#1}{\textnormal{argmax}} \, }}

\newcommand{\ind}{\perp\!\!\!\!\perp} 
\newcommand{\notind}{\not\!\perp\!\!\!\!\perp}

\newcommand{\revision}[1]{{\color{black}{#1}}}
\newcommand{\revisiont}[1]{{\color{black}{#1}}}

\newcommand{\bs}[1]{\mathbf{#1}}
\newtheorem{Assump}{\underline{\bf Assumptions}}

\newtheorem{Definition}{\underline{\bf Definition}}
\usepackage{enumitem}

\begin{document}

\begin{frontmatter}

\title{Estimating Causal Effects with Hidden Confounding using Instrumental Variables and Environments}
\runtitle{Instrumental Variables and Environments}

\begin{aug}
\author{\fnms{James P.} \snm{Long}\ead[label=e1]{jplong@mdanderson.org}}
\address{Department of Biostatistics\\
University of Texas MD Anderson Cancer Center\\
\printead{e1}}

\author{\fnms{Hongxu} \snm{Zhu}}
\address{Department of Biostatistics\\
University of Texas, School of Public Health}

\author{\fnms{Kim-Anh} \snm{Do}}
\address{Department of Biostatistics\\
University of Texas MD Anderson Cancer Center}

\author{\fnms{Min Jin} \snm{Ha}}
\address{Department of Biostatistics\\
Graduate School of Public Health, Yonsei University}

\runauthor{J.P. Long et al.}

\end{aug}

\begin{abstract}
Recent works have proposed regression models which are invariant across data collection environments \citep{rothenhausler2019,peters2016causal,heinze2018invariant,meinshausen2018causality,gimenez2021causal}. These estimators often have a causal interpretation under conditions on the environments and type of invariance imposed. One recent example, the Causal Dantzig (CD), is consistent under hidden confounding and represents an alternative to classical instrumental variable estimators such as Two Stage Least Squares (TSLS). In this work we derive the CD as a generalized method of moments (GMM) estimator. The GMM representation leads to several practical results, including 1) creation of the Generalized Causal Dantzig (GCD) estimator which can be applied to problems with continuous environments where the CD cannot be fit 2) a Hybrid (GCD-TSLS combination) estimator which has properties superior to GCD or TSLS alone 3) straightforward asymptotic results for all methods using GMM theory. We compare the CD, GCD, TSLS, and Hybrid estimators in simulations and an application to a Flow Cytometry data set. The newly proposed GCD and Hybrid estimators have superior performance to existing methods in many settings.
\end{abstract}


\tableofcontents
\end{frontmatter}

\hypertarget{introduction}{%
\section{Introduction}\label{introduction}}

Causal inference is challenging because of confounding and reverse
causality. One solution is to make strong assumptions
about confounding (e.g.~no hidden confounding) and the direction of
causation (e.g. $X$ is a cause of $Y$ and not the reverse). Under these
assumptions, causal parameters may be identifiable from observational
data.

When these assumptions are not valid, instrumental variables (IV) are a
classical method for identifying causal effects. The variable $E$
is an instrument for the $X \rightarrow Y$ causal relation if 1) $E$ is uncorrelated with
the error term in the $Y$ on $X$ regression and 2) $E$ is correlated with $X$ (valid first stage).
IV estimators such as Two Stage Least Squares
(TSLS) remain consistent under hidden confounding and unknown direction
of causality. IV methods date
back to \cite{wright1928tariff} and have more recently been generalized to high dimensional problems \citep{lin2015regularization,gold2020inference,belloni2012sparse} and causal
discovery applications where $X \in \mathbb{R}^p$ is a vector and identifying the
causes of each $X_i$ is of interest \citep{chen2018two}.

Recently, several works have proposed causal estimators based on the concept of data collection \textit{environment} \citep{rothenhausler2019,peters2016causal,heinze2018invariant,meinshausen2018causality,gimenez2021causal}. \cite{peters2016causal} introduced the concept of data collection environment and developed a causal estimator, Invariant Causal Prediction (ICP). In this framework, each observation is collected in an environment. The environment may represent randomized experiments on some of the exposures of interest, shift, and/or noise interventions. Environments are typically discrete and often small in number (e.g.~2 or 3). 

Estimators are constructed from environments based on the principle that parameters in a causal regression model $Y$ on $X$ should be \textit{invariant} across environments while parameters in a merely associational model will vary. As a simple heuristic example, suppose we are interested in estimating the causal effect of $X$ on $Y$. In truth $Y$ is a cause of $X$ and the true causal effect of $X$ on $Y$ is $0$. Standard regression based estimators are inconsistent. More generally with purely observational data it will be impossible to determine the causal effect. However if we have data from two environments, e.g.~an observational environment and an interventional environment where noise is added to $X$, then it is possible (under some conditions) to infer the causal effect as $0$ by noting that the distribution of $Y$ is identical in the observational and interventional environment. This would not be the case if $X$ has a causal effect on $Y$.

The original environment estimator ICP has been generalized to problems with sequential data and non--linear models \citep{pfister2019invariant,heinze2018invariant}. \cite{rothenhausler2019} proposed the Causal Dantzig (CD) environment estimator to address two weaknesses of ICP: computational complexity and inconsistency under hidden confounding. While ICP requires fitting models on all subsets of the exposure variables (making the algorithm superexponential in the number of exposures), the CD estimator has computational burden similar to linear regression. Further, like instrumental variable estimators, the CD is consistent when hidden variables confound the $X \rightarrow Y$ causal relation.

\revision{In this work, we show that the Causal Dantzig can be represented as a generalized method of moments estimator (GMM). This immediately leads to a new estimator, termed the Generalized Causal Dantzig (GCD), which is equivalent to the CD in the two environment case but can be applied with continuous environments, a setting not handled by the original CD. The GMM representation facilitates straightforward asymptotic results based on GMM theory. GMM theory shows how to optimally weight the moment constraints in over--identified problems, which occurs whenever there are more than two data collection environments. Finally, the GMM representation of the CD facilitates construction of a Hybrid estimator which uses both the CD and IV moment constraints for estimating parameters. This Hybrid estimator is consistent in some settings where neither the CD nor TSLS are consistent.} 

This work is organized as follows. In Section \ref{iv-cd} we review the GMM representation of IV estimators and the environment invariance representation of the Causal Dantzig. In Section \ref{gcd} we propose the Generalized Causal Dantzig (GCD), a GMM estimator, and construct a Hybrid estimator which use both IV and CD/GCD moment constraints. Section \ref{asymptotic} derives asymptotic results for the GCD based on the GMM representation of the estimator. In Section \ref{scm}, we assess consistency of the estimators in causal Structural Equation Models. Section \ref{sim} contains simulations which demonstrate some of the potential applications of the GCD and hybrid GCD-IV estimators. In Section \ref{application} we apply IV, CD, GCD, and Hybrid estimators to Flow Cytometry data of \cite{sachs2005causal}. In several cases, we show that IV and Hybrid  estimators identify more plausible causal relations than the CD alone. We conclude with a discussion in Section \ref{discussion}. All code and data for reproducing the computational aspects of this work is available.\footnote{\url{https://github.com/longjp/gcd-code}}

\hypertarget{iv-cd}{%
\section{Instrumental Variables and the Causal Dantzig}\label{iv-cd}}

Let $X \in \mathbb{R}^p$ be a set of \revision{endogenous} exposures and $Y \in \mathbb{R}^1$ be a response. The goal is to estimate the causal effect of $X$ on $Y$. Consider a linear model of the form
\begin{equation}
\label{eq:cm}
Y = X^T\beta_0 + \delta_Y
\end{equation}
where $\mathbb{E}[\delta_Y] = 0$. Under a potential outcomes \citep{angrist1996identification} or a structural equation modelling \citep{pearl2009causal} framework, $\beta_{0j}$ ($j^{th}$ element of $\beta_0$) can be given a causal interpretation as the average treatment effect (ATE) of $X_j$ on $Y$ when shifting $X_j$ by $1$ unit. In either of these frameworks, correlation between $X$ and $\delta_Y$ is induced by hidden confounders which exert a causal effect on $X$ and $Y$. Straightforward regression of $Y$ on $X$ may result in inconsistent estimates of $\beta_0$ when the error term $\delta_Y$ is correlated with $X$.

\revision{In this section we review two approaches to constructing consistent estimators with hidden confounding and reverse causality: the classical Two Stage Least Squares (TSLS) which uses Instrumental variables (IV) and the recently proposed Causal Dantzig which uses data collection environments.}

\hypertarget{iv}{%
\subsection{Instrumental Variable Estimators}\label{iv}}

Instrumental variables techniques, dating back to \cite{wright1928tariff}, use instrumental variables (IVs) $E \in \mathbb{R}^q$ to construct consistent estimates of $\beta$ in the presence of hidden confounding. Suppose that 1) the instruments $E$ are uncorrelated with the error term $\delta_Y$ ($\mathbb{E}[E\delta_Y] = 0$) and 2) $\mathbb{E}[EX^T] \in \mathbb{R}^{q \times p}$ is of rank at least $p$. The latter condition implies that $q \geq p$ (i.e. there are at least as many instruments as exposures). We review the construction of IV estimators from a GMM perspective. See \cite{matyas1999generalized} for more background.

Let $Z = (Y,X,E)$ and $g_{IV}(Z,\beta) = E(Y-X^T\beta)$. Then the true causal parameter $\beta_0$ is the unique solution to
\begin{equation}
\label{eq:tsls-gmm}
m_{IV}(\beta) = \mathbb{E}[g_{IV}(Z,\beta)] = \mathbb{E}[E(Y-X^T\beta)] = 0.
\end{equation}
This can be seen by noting
\begin{equation*}
\mathbb{E}[E(Y-X^T\beta)] = \underbrace{\mathbb{E}[E\delta_Y]}_{=0} + \underbrace{\mathbb{E}[EX^T]}_{\text{rank } \geq p}(\beta_0 - \beta).
\end{equation*}
The rank condition implies that the null-space of the matrix is $0$ implying $\beta_0$ is the only solution. To construct a consistent estimator, the expectation is approximated with a sample. Define \(\textbf{X} \in \mathbb{R}^{n\times p}\),
\(\textbf{E} \in \mathbb{R}^{n \times q}\),
\(\textbf{Y} \in \mathbb{R}^{n \times 1}\) to be a matrices of \(n\)
i.i.d. observations. With $i$ indexing observations, we have
\begin{equation}
\label{eq:gmm-tsls}
\widehat{m}_{IV}(\beta) = \frac{1}{n}\sum_{i=1}^n E_i (Y_i - X_i^T\beta) = \frac{1}{n}\bs{E}^T(\bs{Y}-\bs{X}\beta).
\end{equation}
When $q > p$, the model is over--identified and there will typically be no $\widehat{\beta}$ such that $\widehat{m}_{IV}(\widehat{\beta}) = 0$ in Equation \eqref{eq:gmm-tsls}. In this case, the standard GMM approach is to use estimator
\begin{equation}
\label{eq:gmm-tsls-est}
\widehat{\beta}_{IV}(\widehat{W}) = \argmin{\beta} ||\widehat{m}_{IV}(\beta)||^2_{\widehat{W}} = \argmin{\beta} \widehat{m}_{IV}(\beta)^T \widehat{W} \widehat{m}_{IV}(\beta)
\end{equation}
where $\widehat{W} \succ 0$ is a positive definite weighting matrix. The TSLS IV estimator uses
\begin{equation*}
\widetilde{W} = \left(\frac{1}{n}\bs{E}^T\bs{E}\right)^{-1}.
\end{equation*}
This weight matrix is chosen for asymptotic efficiency considerations which we discuss further in Section \ref{an} (see also Section 1.3.4.2 of \cite{matyas1999generalized}). With $\widetilde{W}$, Equation \eqref{eq:gmm-tsls-est} has the form
\begin{equation}
\label{eq:tsls}
\widehat{\beta}_{TSLS} \equiv \widehat{\beta}_{IV}(\widetilde{W}) = (\widehat{\bs{X}}^T \widehat{\bs{X}})^{-1}\widehat{\bs{X}}\bs{Y}
\end{equation}
where $\widehat{\bs{X}} = \bs{E}(\bs{E}^T \bs{E})^{-1}\bs{E}^T\bs{X}$. The TSLS estimator derives its name from the fact that it is computed by first regressing $X$ on $E$ (first stage) and then regressing $Y$ on the predicted values from the first stage (second stage).

Note that when $p=q$ (just identified case) the unique $\widehat{\beta}_{IV}(\widehat{W})$ does not depend on $\widehat{W}$ and has the form
\begin{equation*}
\widehat{\beta}_{IV}(\widehat{W}) = (\bs{E}^T \bs{X})^{-1}\bs{E}^T\bs{Y}.
\end{equation*}

\hypertarget{causal-dantzig}{%
\subsection{Causal Dantzig}\label{causal-dantzig}}

The Causal Dantzig (CD) uses environments to estimate $\beta_0$ in Equation \eqref{eq:cm}. Each observation belongs to one of a discrete set of environments.
Let $\mathcal{E} = (1,2,\ldots)$ be a set of data collection environments with $\# \mathcal{E}$ denoting the number of environments (at least $2$). Let $X^e$ ($Y^e$) denote exposures (response) collected in environment $e \in \mathcal{E}$. For any $f,g \in \mathcal{E}$, the CD seeks a $\beta$ which satisfies
\begin{equation}
\label{eq:cdinv}
\mathbb{E}[X^f(Y^f - X^{f^T}\beta)] = \mathbb{E}[X^g(Y^g - X^{g^T}\beta)].
\end{equation}
\revision{Under conditions specified in Section \ref{asymptotic}, Equation \eqref{eq:cdinv} will have a unique solution which equals the causal estimand $\beta_0$.} The CD estimator is constructed by enforcing sample based versions of constraints in Equation \eqref{eq:cdinv}. Since $X^e \in \mathbb{R}^p$, when $\# \mathcal{E} = 2$ (two data collection environments), the invariances specified in Equation \eqref{eq:cdinv} produce $p$ sample constraints on $\beta \in \mathbb{R}^p$. Suppose there are $n_e$ observations from environment $e$. Let $\bs{X}^e \in$ $\mathbb{R}^{n_e \times p}$ ($\bs{Y}^e \in$ $\mathbb{R}^{n_e}$) represent the design matrix (response vector) for environment $e$. Then the sample version of constraints in Equation \eqref{eq:cdinv} (with $f=2$ and $g=1$) is
\begin{equation*}
\frac{1}{n_2}\bs{X}^{2^T}(\bs{Y}^2 - \bs{X}^2\widehat{\beta}_{CD}) = { \frac{1}{n_1} } \bs{X}^{1^T}(\bs{Y}^1 - \bs{X}^1\widehat{\beta}_{CD}).
\end{equation*}
Solving for $\widehat{\beta}_{CD}$ one obtains
\begin{equation}
\label{eq:cd}
\widehat{\beta}_{CD} = \left(\frac{1}{n_2}\bs{X}^{2^T}\bs{X}^2 - \frac{1}{n_1} \bs{X}^{1^T}\bs{X}^1\right)^{-1} \left(\frac{1}{n_2}\bs{X}^{2^T}\bs{Y}^2 - \frac{1}{n_1}\bs{X}^{1^T}\bs{Y}^1\right)
\end{equation}
assuming the inverse exists (see Equation 7 of \cite{rothenhausler2019}). \revision{Equation \eqref{eq:cd} shows that the CD exploits how the covariance structure of $X$ changes with the environment to construct a consistent estimate of $\beta$. This is in contrast to TSLS which exploits how the mean of $X$ changes with the instrument $E$. The CD is a consistent and asymptotically normal estimator of $\beta_0$ under conditions specified in Sections \ref{asymptotic} and \ref{scm}.}

\hypertarget{gcd}{%
\section{New Estimators}\label{gcd}}

\revision{We now construct the Generalized Causal Dantzig (GCD) and a Hybrid estimator. The GCD extends the CD to work in new settings (e.g. continuous environments). The Hybrid estimator imposes both TSLS and CD moment constraints to produce an estimator with potentially better asymptotic properties than either the CD or TSLS. Finally we discuss connections between the GCD and estimators proposed by Lewbel \cite{lewbel2012using}.}

\subsection{Generalized Causal Dantzig}

Following notation used for the IV estimators in Section \ref{iv} where $Z=(Y,X,E)$, define the Generalized Causal Dantzig (GCD) using the moment conditions
\begin{equation}
\label{eq:cd-gmm-const}
m_{GCD}(\beta) = \mathbb{E}[g_{GCD}(Z,\beta)] = \mathbb{E}[vec(EX^T)(Y-X^T\beta)] = 0.
\end{equation}
\revision{Here $EX^T \in \mathbb{R}^{q \times p}$ and $vec(EX^T)$ vectorizes (column stacks) the matrix $EX^T$ \citep{henderson1981vec}. Specifically
\begin{equation*}
vec(EX^T) = vec\begin{pmatrix}
E_1X_1 & \ldots & E_1X_p\\
\vdots & \vdots & \vdots\\
E_qX_1 & \ldots & E_qX_p
\end{pmatrix} = \begin{pmatrix}
E_1X_1\\
\vdots\\
E_qX_1\\
\vdots\\
E_1X_p\\
\vdots\\
E_qX_p
\end{pmatrix} \in \mathbb{R}^{qp}.
\end{equation*}}
We now justify the name Generalized Causal Dantzig (GCD) by showing an equivalence between GCD moment conditions in Equation \eqref{eq:cd-gmm-const} and CD invariance constraints in Equation \eqref{eq:cdinv}.\revisiont{ To achieve this result, we first translate CD environment data notation ($(Y^e,X^e)$ for $e \in \mathcal{E}$) to IV and GCD data notation ($Z=(Y,X,E)$). We term this translation $Z$--Encoding.
\begin{Definition}[$Z$--Encoding]
\label{set:code2}
Let $(Y^e,X^e)$ denote data collected in environment $e \in \mathcal{E}$. Define $Y=Y^e$, $X=X^e$ and $R=e$. Further encode the categorical variable $R$ with random variable $E \in \mathbb{R}^{\# \mathcal{E}-1}$ where
\begin{equation}
\label{eq:iv-env-const}
E_f = \begin{cases} 
      s_{f1} & R=f \\
      s_{f0} & R \neq f
     \end{cases}
\end{equation}
for $f \in \{1,\ldots,\#\mathcal{E}-1\}$. Select $s_{f1}$ and $s_{f0}$ such that $\mathbb{E}[E_f]=0$. The $Z$--Encoding of environment data $(Y^e,X^e)$ is $(Y,X,E)$.
\end{Definition}
\begin{thm} \label{thm:cd-gmm}
Let $(Y^e,X^e)$ for $e \in \mathcal{E}$ be environment data and $Z=(Y,X,E)$ be its $Z$--Encoding constructed according to Definition \ref{set:code2}. Then the Causal Dantzig invariance constraints (Equation \eqref{eq:cdinv}) applied to the environment data and the GCD moment constraints (Equation \eqref{eq:cd-gmm-const}) applied to the $Z$--Encoding are identical. Specifically
\begin{equation*}
\{\beta : m_{GCD}(\beta)=0\} = \{\beta : \mathbb{E}[X^f (Y^f - X^{f^T}\beta)] = \mathbb{E}[X^g (Y^g - X^{g^T}\beta)] \, \, \forall f,g \in \mathcal{E}\}.
\end{equation*}
\end{thm}}
See Section \ref{cd-gmm-proof} for a proof. The result has conceptual and practical implications. On the conceptual side, comparing the IV moment constraints (Equation \eqref{eq:tsls-gmm}) with the GCD moments constraints (Equation \eqref{eq:cd-gmm-const}) shows the environment and instrument ($E$) play similar roles in the estimators. TSLS requires each instrument/environment to be orthogonal to $Y-X^T\beta$ while CD requires each instrument/environment be orthogonal to each element of $X(Y-X^T\beta)$. On a practical side, the GCD provides a natural generalization of the CD to problems with continuous environments/instruments and GMM theory enables straightforward derivation of GCD asymptotics. This includes optimal weighting of invariance (equivalently moment) criteria in the overidentified case using a two--step estimator. We explore these ideas further in Sections \ref{asymptotic} and \ref{sim}.

We now construct the GCD estimator by imposing the constraints in Equation \eqref{eq:cd-gmm-const} on the sample. Note that $vec(EX^T) \in \mathbb{R}^{qp}$ induces $qp$ constraints on $\beta$. Define
\begin{align*}
\bs{E} \bullet \bs{X} &= \begin{pmatrix}
vec(E_1X_1^T)^T\\
\vdots \\
vec(E_nX_n^T)^T\\
\end{pmatrix}\\
&=
\begin{pmatrix}
E_{11}X_{11} & \dots & E_{1q}X_{11} & \dots & E_{11}X_{1p} & \dots & E_{1q}X_{1p}\\
\vdots & \vdots & \vdots & \vdots & \vdots & \vdots & \vdots \\
E_{n1}X_{n1} & \dots & E_{nq}X_{n1} & \dots & E_{n1}X_{np} & \dots & E_{nq}X_{np}
\end{pmatrix} \\
&\in \mathbb{R}^{n \times pq}.
\end{align*}
Define the GCD moment equations as
\begin{equation}
\label{eq:cd-gmm-sample}
\widehat{m}_{GCD}(\beta) = \frac{1}{n}(\bs{E}\bullet\bs{X})^T(\bs{Y}-\bs{X}\beta).
\end{equation}
and the resulting GCD estimator as
\begin{equation}
\label{eq:cd-gmm-est}
\widehat{\beta}_{GCD}(\widehat{W}) = \argmin{\beta} ||\widehat{m}_{GCD}(\beta)||^2_{\widehat{W}}
\end{equation}
where $\widehat{W} \succ 0$ is some weighting matrix. When $E$ is constructed from two environments, its dimension is $1$ because $q = \# \mathcal{E} -1=1$. In this case the $qp = p$ constraints just identify $\beta$. Further $\widehat{\beta}_{GCD}(\widehat{W})$ is invariant to different choices of $\widehat{W}$ and identical to the original Causal Dantzig estimator $\widehat{\beta}_{CD}$. The following theorem formalizes these results.

\begin{thm} \label{thm:cd-gmm-est}
Suppose \revisiont{$\widehat{W} \succ 0$ and} $\bs{X}^T(\bs{E}\bullet \bs{X}) \in \mathbb{R}^{p \times pq}$ has column rank $p$. Then
\begin{enumerate}
\item The \revisiont{unique} minimizer in Equation \eqref{eq:cd-gmm-est} is
\begin{equation*}
\widehat{\beta}_{GCD}(\widehat{W}) = (\bs{X}^T(\bs{E}\bullet\bs{X})\widehat{W}(\bs{E}\bullet\bs{X})^T\bs{X})^{-1}(\bs{X}^T(\bs{E}\bullet \bs{X})\widehat{W}(\bs{E} \bullet \bs{X})^T \bs{Y}).
\end{equation*}
\item If $q = dim(E) = 1$ (just identified case), then $\widehat{\beta}_{GCD}(\widehat{W})$ is invariant to the choice of $\widehat{W}$ and 
\begin{equation*}
\widehat{\beta}_{GCD}(\widehat{W}) = \widehat{\beta}_{GCD} = ((\bs{E}\bullet \bs{X})^T \bs{X})^{-1}(\bs{E} \bullet \bs{X})^T\bs{Y}.
\end{equation*}
\item \revisiont{If $Z=(Y,X,E)$ is the $Z$--Encoding (Definition \ref{set:code2}) of environment data $(Y^e,X^e)$ with $\# \mathcal{E}=2$ (two environments), then $\widehat{\beta}_{GCD} = \widehat{\beta}_{CD}$.}
\end{enumerate}
\end{thm}

See Section \ref{cd-gmm-est-proof} for a proof. With more than two environments, $\widehat{\beta}_{GCD}(\widehat{W})$ will generally not be equivalent to the $\# \mathcal{E} > 2$ CD estimators proposed in \cite{rothenhausler2019}. \revision{\cite{rothenhausler2019} proposed two approaches for adapting the CD to the case with more than two environments: 1) Merge environments to obtain two distinct environments. Fit the two environment CD estimator on the resulting data. 2) Fit a minimax estimator which seeks to satisfy the CD invariance constraints in a one--versus--all environment approach (Equation 8 in \citep{rothenhausler2019}).  No theory is given to guide merging of environments (if approach 1 is taken) or how to compute uncertainties on parameter estimates (if approach 2 is taken). The CD was previously applied to the Flow Cytometry Application studied in Section 6 of this work \citep{meinshausen2016methods}. In that application, a third strategy was used for adapting the CD to more than two environments which involves iteratively fitting the two environment CD to pairs of environments (see Section 6.2 of this work and \cite{meinshausen2016methods} for a more detailed description). In contrast, the GCD with more than two environments uses the matrix $\widehat{W}$ to weight moment conditions. GMM theory is then used to derive the optimal weight matrix (minimizes asymptotic variance) which can be estimated using standard two--step procedures. GMM theory provides (asymptotically) valid uncertainty estimates. This approach is used both in the simulations and the Flow Cytometry application of Section 6.} We discuss optimal selection of $\widehat{W}$ in Section \ref{an}.

\revision{
\subsection{Hybrid Estimator}
\label{sec:hybrid}

The GMM representation of the GCD in Equation \ref{eq:cd-gmm-const} and the GMM representation of the TSLS estimator in Equation \ref{eq:tsls-gmm} share a similar structure. The GCD enforces orthogonality between $Y-X^T\beta$ and $vec(EX^T)$ while TSLS enforces orthogonality between $Y-X^T\beta$ and $E$. Using GMMs, it is straightforward to construct estimators which enforce both of these constraints. We define the Hybrid estimator moment conditions
\begin{equation}
\label{eq:hybrid}
m_H(\beta) = \mathbb{E}\left[\begin{pmatrix} E \\ vec(EX^T) \end{pmatrix}(Y-X^T\beta)\right] = 0.
\end{equation}
This estimator enforces orthogonality between $Y-X^T\beta$ and both $vec(EX^T)$ and $E$, resulting in a total of $q + qp$ constraints. Following standard GMM practice, the sample version of these orthogonality constraints is
\begin{equation*}
\widehat{m}_H(\beta) = \frac{1}{n} (\bs{E}, \bs{E}\bullet\bs{X})^T(\bs{Y}-\bs{X}\beta)
\end{equation*}
where $(\bs{E}, \bs{E}\bullet\bs{X}) \in \mathbb{R}^{n \times (q + qp)}$ columns joins the matrices $\bs{E}$ and $\bs{E}\bullet\bs{X}$. The hybrid estimator is
\begin{equation}
\label{eq:hybrid-est}
\widehat{\beta}_{H}(\widehat{W}) = \argmin{\beta} ||\widehat{m}_{H}(\beta)||^2_{\widehat{W}}
\end{equation}
where $\widehat{W} \in \mathbb{R}^{(q+qp) \times (q+qp)} \succ 0$ is some weighting matrix. Since $\widehat{\beta}_{IV}$, $\widehat{\beta}_{GCD}$, and $\widehat{\beta}_{H}$ are all GMMs, asymptotic results for these estimators can be derived using GMM theory. We do this in the following section.

\subsection{Relation to Lewbel \cite{lewbel2012using}}

Lewbel \cite{lewbel2012using} constructed consistent estimators of causal effects with hidden confounding by exploiting heteroscedasticity in endogenous variables. The relationship between Lewbel's estimators and the CD was briefly discussed in the original CD paper \cite{rothenhausler2019}. \cite{rothenhausler2019} claimed that the methods are different because Lewbel uses exogenous variables while the CD directly exploits endogenous covariance structure, ``resulting in a different method.'' Complicating comparison of the methods is the fact that the original CD was constructed based on invariance and environments (Equation \eqref{eq:cdinv}) while Lewbel used GMM.

Our result that the CD can be represented as a GMM (Theorem \eqref{thm:cd-gmm-est}), facilitates comparison between the methods. In fact, for certain choices of Lewbel's variable $Z$, the CD and Lewbel are identical. See Supplementary Material Section \ref{lewbel} for details. Neither Lewbel's estimators nor the CD/GCD/Hybrid completely overlap in the cases they consider. Lewbel considers fully simultaneous systems and models additional exogenous variable which are not part of this work. Lewbel primarily considers univariate endogenous variable models, briefly developing an extension of the GMM to the $p=2$ case (Section 3.3). Here we study the $p>1$ case in depth. The Hybrid estimator is to our knowledge completely new and not considered in Lewbel.

}

\hypertarget{asymptotic}{%
\section{Asymptotic Properties of Estimators}\label{asymptotic}}

The GMM representation of the IV, GCD, and Hybrid estimators makes derivation of asymptotic properties straightforward. We first discuss consistency and then asymptotic normality.

\hypertarget{consistency}{%
\subsection{Consistency}\label{consistency}}

Assumptions \ref{assump:consist} and Theorem \ref{thm:consist_gen} below may be specialized to the IV, GCD, or Hybrid estimators by substituting in the corresponding $m$, $\widehat{m}$, and $\widehat{\beta}(\widehat{W})$. For example, for the GCD, let $m=m_{GCD}$, $\widehat{m}=\widehat{m}_{GCD}$, and $\widehat{\beta}(\widehat{W}) = \widehat{\beta}_{GCD}(\widehat{W})$.

\begin{Assump} \label{assump:consist}
Suppose
\begin{enumerate}[label=(\alph*)]
\item $m(\beta)$ exists and is finite for all $\beta \in \Theta$
\item $m(\beta) = 0$ iff $\beta=\beta_0$, the causal parameter
\item $\widehat{m}(\beta) \rightarrow_p m(\beta)$ uniformly in $\beta$
\item $\widehat{W} \rightarrow_P W$ where $\widehat{W},W \succ 0$
\end{enumerate}
\end{Assump}
\revisiont{
\begin{thm} \label{thm:consist_gen}
Under Assumptions \ref{assump:consist}, $\widehat{\beta}(\widehat{W})$ is a consistent estimator of $\beta_0$.
\end{thm}}
See Theorem 1.1 and Section 1.3.4.1 of \cite{matyas1999generalized} for a proof and application to the i.i.d. case. \revision{Assumption \ref{assump:consist} d) is the simplest to satisfy. For example, $\widehat{W} = W = I$ satisfies the assumption. In general $\widehat{W}$ will be chosen to be a consistent estimator of a $W$ which minimizes the asymptotic variance of the estimator, as discussed in Section \ref{an}. Assumption \ref{assump:consist} a) will generally hold whenever error terms have sufficient moments. While Assumption \ref{assump:consist} c) is strong, it may be replaced with assumptions restricting $\beta$ to a compact subset of $\mathbb{R}^p$ and continuity of $m(\beta)$ \citep{hansen1982large,newey1994large}. Assumption \ref{assump:consist} b) is closely related to the concepts of instrument/environment validity. We give conditions under which Causal Structural Equation Models (SEM) will satisfy Assumption \ref{assump:consist} b) in Section \ref{consist}. These results clarify the strengths and weaknesses of the IV, GCD, and Hybrid estimators.}

\hypertarget{an}{%
\subsection{Asymptotic Normality}\label{an}}

Asymptotic normality of the GCD and Hybrid estimators can be derived from standard GMM theory. The asymptotic theory provides guidance on selection of the weight matrix $\widehat{W}$ in the overidentified case (recall that overidentification will occur for the GCD whenever there are more than two environments or more generally when the dimension of $E$ is greater than $1$). The optimal weight matrix is chosen to minimize the asymptotic variance. For the GCD, this optimal weight matrix can be estimated using a GMM two--step procedure.

Assumptions \ref{assump:an} below are used for showing asymptotic normality of GMM estimators. These may be specialized to the IV, GCD, or Hybrid estimators by substituting in the corresponding $g$, $m$, $\widehat{m}$, and $\widehat{\beta}$. For example, for the GCD, let $g=g_{GCD}$, $m=m_{GCD}$, $\widehat{m}=\widehat{m}_{GCD}$, and $\widehat{\beta}(\widehat{W})=\widehat{\beta}_{GCD}(\widehat{W})$.

\begin{Assump} \label{assump:an}
Suppose
\begin{enumerate}[label=(\alph*)]
\item $g(Z,\beta)$ is continuously differentiable for $\beta \in \Theta$.
\item Let $k=dim(\widehat{m})$. Define
\begin{equation*}
\widehat{M}(\beta) = \frac{\partial \widehat{m}(\beta)}{\partial \beta} \in \mathbb{R}^{k \times p}
\end{equation*}
and suppose $\widehat{\beta} \rightarrow_P \beta_0$. Suppose there exists an $M \in \mathbb{R}^{k \times p}$ of full column rank such that
\begin{equation*}
\widehat{M}(\widehat{\beta}) \rightarrow_P M.
\end{equation*}
\item $V \equiv Var(g(Z,\beta_0))$ exists (i.e. $g(Z,\beta_0)$ has two moments).
\end{enumerate}
\end{Assump}
\revisiont{
\begin{thm} \label{thm:an}
Under Assumptions \ref{assump:consist} and \ref{assump:an}, 
\begin{equation*}
\sqrt{n}(\widehat{\beta}(\widehat{W}) - \beta_0) \rightarrow_d N(0,\Sigma(W))
\end{equation*} with asymptotic variance
\begin{equation*}
\Sigma(W) = (M^TWM)^{-1}M^TWVWM(M^TWM)^{-1}.
\end{equation*}
\end{thm}}
See Section 1.3.4.1 of \cite{matyas1999generalized} for these results. The asymptotic variance is minimized by setting $W=\bar{W}\propto V^{-1}$ which results in
\begin{equation}
\label{eq:avar}
\Sigma(\bar{W}) = (M^TV^{-1}M)^{-1}.
\end{equation}

We now review weighting for the TSLS estimator. For the instrumental variables estimator
\begin{align*}
\bar{W} &\propto \mathbb{E}[g_{IV}(Z,\beta_0)g_{IV}(Z,\beta_0)^T]^{-1}\\
&= \mathbb{E}[E(Y-X^T\beta_0)(Y-X^T\beta_0)E^T]^{-1}\\
&= (\mathbb{E}[EE^T\delta_Y^2])^{-1}.
\end{align*}
When $E$ is independent of $\delta_Y$, $\bar{W} \propto \mathbb{E}[EE^T]^{-1}$. The matrix  $\widetilde{W} = (n^{-1}\bs{E}^T\bs{E})^{-1}$ is a consistent estimator of $\bar{W}$ and thus is an asymptotically optimal weighting for the IV estimator. The estimator $\widehat{\beta}_{IV}(\widetilde{W})$ can be computed using a set of two regressions, giving it the name Two Stage Least Squares. Since consistency of $\widehat{\beta}_{IV}$ only requires orthogonality of $E$ and $\delta_Y$, there may be cases where $\widehat{\beta}_{IV}(W)$ is a consistent, asymptotically normal estimate of $\beta$ but $W = \mathbb{E}[EE^T]^{-1}$ is not asymptotically efficient. In these cases, an asymptotically efficient IV estimator can be constructed assuming one can construct a consistent estimate of $\bar{W}$. One possibility is to use two--step procedures. First a pilot estimator $\widehat{\beta}_{IV}(W)$ is computed using some initial weight matrix $W$. Possible choices include $W=I$ (identity) or $W=\widetilde{W}$ (TSLS weight). Residuals are defined as
\begin{equation*}
\widehat{\delta}_{Y,i} = Y_i - X_i^T\widehat{\beta}_{IV}(W).
\end{equation*}
Define $\widehat{\bs{\Sigma}}$ as a diagonal matrix with $\widehat{\bs{\Sigma}}_{ii} = \widehat{\delta}_{Y,i}^2$ and
\begin{equation*}
\widehat{W}_{TS} = \left(\frac{1}{n} \bs{E}^T\widehat{\bs{\Sigma}}\bs{E}\right)^{-1}.
\end{equation*}
If $\widehat{W}_{TS} \rightarrow_P \bar{W}$, then $\widehat{\beta}_{IV}(\widehat{W}_{TS})$ is asymptotically efficient. This two--step procedure is sometimes referred to as optimal GMM. See Section 6.4.2 of \cite{cameron2005microeconometrics} for a discussion of two--step optimal GMM estimators and comparison with TSLS.

For the GCD, the asymptotically optimal estimators can be constructed from two--step procedures, following the same strategy as used for IV. The following theorem provides formal justification of this approach. Note that this is only necessary in the over--identified case since in the just identified case the estimator is invariant to different choices of $W$.
\begin{thm} \label{thm:cd-gmm-est-an}
Suppose Assumptions \ref{assump:consist} and \ref{assump:an} hold for the GCD estimator. Let $\widehat{W} \succ 0$ be some initial weight matrix such that $\widehat{W} \rightarrow W \succ 0$. Define
\begin{equation*}
\widehat{\delta}_{Y,i} = Y_i - X_i^T\widehat{\beta}_{GCD}(\widehat{W}).
\end{equation*}
Let $\widehat{\bs{\Sigma}}$ be a diagonal matrix with $\widehat{\bs{\Sigma}}_{ii} = \widehat{\delta}_{Y,i}^2$ and define
\begin{equation*}
\widehat{W}_{GCD} = \left(\frac{1}{n} (\bs{E \bullet X})^T\widehat{\bs{\Sigma}}(\bs{E \bullet X})\right)^{-1}.
\end{equation*}
Then $\widehat{\beta}_{GCD}(\widehat{W}_{GCD})$ is asymptotically efficient with variance specified in Equation \eqref{eq:avar}.
\end{thm}
See Section \ref{cd-gmm-est-an-proof} for a proof. For the GCD, the asymptotic variance can be estimated by first estimating $M$ with
\begin{equation*}
\widehat{M} = \frac{1}{n}(\bs{E} \bullet \bs{X})^T \bs{X}
\end{equation*}
and $V$ with
\begin{equation*}
\widehat{V} = \frac{1}{n}(\bs{E} \bullet \bs{X})^T \bs{\widehat{\Sigma}} (\bs{E} \bullet \bs{X}).
\end{equation*}
Then
\begin{equation}
\label{eq:av-est}
\widehat{\Sigma}(W) = \left(\widehat{M}^T\widehat{V}^{-1}\widehat{M}\right)^{-1}.
\end{equation}

\revision{
\hypertarget{scm}{%
\section{Causal Models and Consistency}\label{scm}}

\revisiont{
Recall that Assumption \ref{assump:consist} b) states
\begin{equation*}
m(\beta) = 0 \text{ iff } \beta=\beta_0.
 \end{equation*}
We now consider Causal Structural Equation Models (SEM) which guarantee that Assumption \ref{assump:consist} b) holds. In combination with regularity Assumptions \ref{assump:consist} a), c), and d), these will ensure consistency of the estimators by Theorem \ref{thm:consist_gen}. We also contrast these estimators and assumptions with Independent Instrumental Variable (IIV) methods.
\begin{Assump}[Causal SEM] \label{assump:causal-sem}
Let $Z=(Y,X,E)$ be generated from a Causal SEM with independent exogenous variables $\epsilon_E,\epsilon_H,\epsilon_X,\epsilon_Y$ and endogenous variables $X \in \mathbb{R}^p$ and $Y \in \mathbb{R}^1$:
\begin{align*}
E &\leftarrow f_E(\epsilon_E)\\
H &\leftarrow f_H(\epsilon_H) \nonumber \\
X &\leftarrow f_X(H,E,\epsilon_X) = \mathbb{E}[f_X(H,E,\epsilon_X)|E] + \delta_X(H,E,\epsilon_X) \nonumber \\
Y &\leftarrow f_Y(H,X,\epsilon_Y) = X^T\beta_0 + \delta_Y(H,X,\epsilon_Y) \nonumber 
\end{align*}
where
\begin{align*}
\mathbb{E}[Y|do(X=x)] &= x^T\beta_0\\
\delta_Y(H,X,\epsilon_Y) &= \delta_Y^1(H,\epsilon_Y) + \delta_Y^2(X,\epsilon_Y)\\
\mathbb{E}[E] &= 0
\end{align*}
\end{Assump}
\begin{figure}
\centering
\includegraphics[width=1\linewidth]{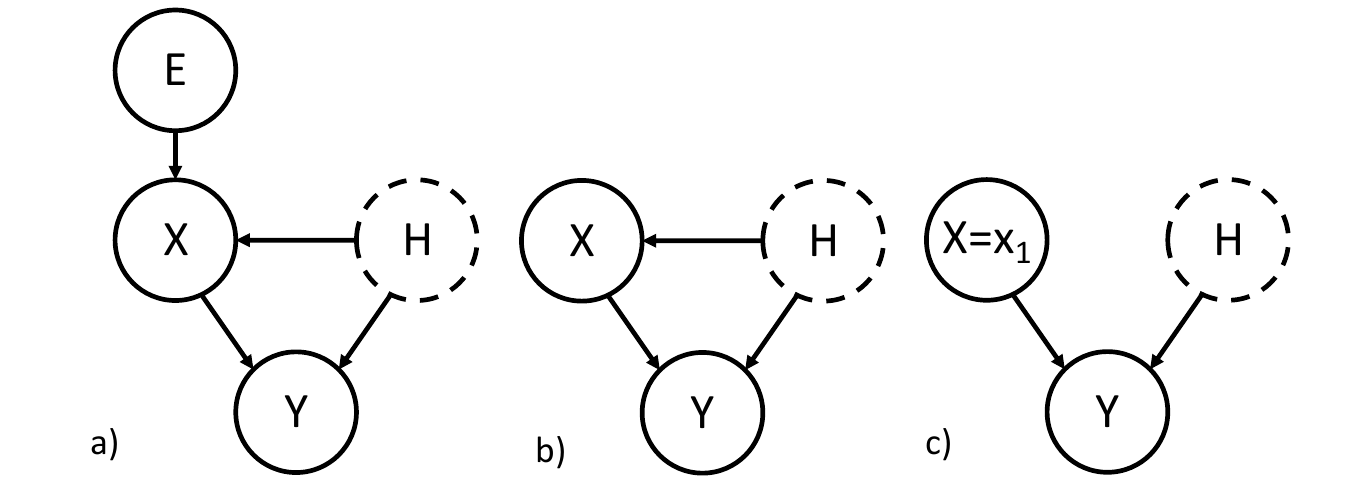}
\caption{a) Causal DAG model with instrument/environment $E$ b) Causal DAG for observational data. c) Causal DAG for interventional data ($do(X=x_1)$) \label{fig:figs2}}
\end{figure}
The corresponding Directed Acyclic Graph (DAG) is show in Figure \ref{fig:figs2} a). The dashed circle around $H$ indicates that it is not measured and thus a hidden confounder. For GCD and Hybrid consistency, we need an additional assumption.
\begin{Assump}[Error Decomposition]
\label{assump:decomp}
\begin{equation*}
\delta_X(H,E,\epsilon_X) = \delta^1_X(E,\epsilon_X) + \delta^2_X(H,\epsilon_X).
\end{equation*}
\end{Assump}

We discuss this decomposition assumption further in Section \ref{sec:decomp}. We have the following result.
}
\revisiont{
\begin{thm}
\label{thm:consist}
Suppose Assumptions \ref{assump:consist} a) c) and d) and Assumption \ref{assump:causal-sem} hold.
\begin{enumerate}
\item Then $m_{IV}(\beta_0)=0$ and $\widehat{\beta}_{IV}(\widehat{W})$ is consistent if $\mathbb{E}[EX^T]$ has column rank $p$
\item If Assumption \ref{assump:decomp} holds, then $m_{GCD}(\beta_0)=0$ and $\widehat{\beta}_{GCD}(\widehat{W})$ is consistent if $\mathbb{E}[vec(EX^T)X^T]$ has column rank $p$.
\item If Assumption \ref{assump:decomp} holds, then $m_{H}(\beta_0)=0$ and $\widehat{\beta}_H(\widehat{W})$ is consistent if $\mathbb{E}\left[\begin{pmatrix} E \\ vec(EX^T) \end{pmatrix}X^T\right]$ has column rank $p$.
\end{enumerate}
\end{thm}}
See Section \ref{cd-gmm-proof} for a proof. Column rank conditions in Theorem \ref{thm:consist} depend on observed random variables $X$ and $E$ and are empirically verifiable. The structure of $f_X$ dictates whether these column rank conditions hold and which estimator (IV, GCD, or Hybrid) will be most appropriate for a given problem. We discuss some specific cases now. For notational convenience let 
\begin{equation*}
r(E) \equiv \mathbb{E}[f_X(H,E,\epsilon_X)|E].
\end{equation*}
\begin{itemize}
\item \underline{IV:} The IV rank condition is $\mathbb{E}[EX^T] = \mathbb{E}[Er(E)^T]$.  The IV estimator leverages how the instruments change the mean of $X$. If $r(E)_j=0$ for any $j \in \mathbb{R}^p$, then IV is not consistent. Further if $q = dim(E) < dim(X) = p$, then the IV is not consistent because the column rank of $\mathbb{E}[EX^T]$ is bounded by $q < p$. Thus with a single instrument, IV may only estimate the causal effect of a single exposure on the response.
\item \underline{GCD:} Suppose $r(E) = 0$ so the IV estimator is not consistent. The GCD rank condition may be rewritten as
\begin{align*}
\mathbb{E}[vec(EX^T)X^T] &= \mathbb{E}[vec(E\delta_X(H,E,\epsilon_X)^T)\delta_X(H,E,\epsilon_X)^T]\\
&= \mathbb{E}[vec(E\delta^1_X(E,\epsilon_X)^T)\delta^1_X(E,\epsilon_X)^T].
\end{align*}
The GCD leverages how $E$ shifts higher moments of $X$ ($\delta^1_X(E,\epsilon_X)$).  The GCD does not require $q \geq p$ because $\mathbb{E}[vec(EX^T)X^T] \in \mathbb{R}^{pq \times p}$. With a single binary instrument/environment taking values $s_1$ or $s_0$, the GCD will be consistent if 
\begin{equation*}
\mathbb{E}[XX^T|E=s_1] - \mathbb{E}[XX^T|E=s_0]
\end{equation*}
is column rank $p$.
\item \underline{Hybrid:} Consistency of the Hybrid estimator is weaker than for the GCD since
\begin{equation*}
\text{colrank}\left(\mathbb{E}\left[\begin{pmatrix} E \\ vec(EX^T) \end{pmatrix}X^T\right]\right) \geq \text{colrank}\left(\mathbb{E}[vec(EX^T)X^T]\right).
\end{equation*}
The Hybrid estimator leverages how $E$ shifts the mean (via the IV constraints) or higher moments (via the GCD constraints) of $X$.
\end{itemize}

\revisiont{Theorem \ref{thm:consist} shows consistency of estimators under assumptions on the data generating model. In some settings, particular elements of the estimator may be consistent while others are not, e.g. $\widehat{\beta}_j(\widehat{W}) \rightarrow_P \beta_{0j}$ for some but not all $j \in \{1, \ldots, p\}$. Such partial identifiability/consistency results have been derived for IV estimators under weaker conditions than considered here \citep{rothenhausler2021anchor,peters2016causal}. Detailed consideration of these cases for the GCD and Hybrid estimators is beyond the scope of this work.}

\subsection{Decomposition Condition and Do Operators}
\label{sec:decomp}

Assumption \ref{assump:decomp} will hold for many models including an additive hidden confounder model where $\delta_X^2(H,\epsilon_X) = \gamma(H)$ for some function $\gamma$. The assumption will not hold when the instrument/environment $E$ represents a do operator. Consider the structural model for $X$
\begin{equation*}
\label{eq:do-operator}
X \gets f_X(H,E,\epsilon_X) = \begin{cases} 
      f_X(H,\epsilon_X) &\text{ if } E=-1\\
      x_1 &\text{ if } E=1
     \end{cases}.\\
\end{equation*}
for some $f_X(H,\epsilon_X)$. When $E=-1$, the data is observational and generated from the DAG in Figure \ref{fig:figs2} b). When $E=1$, the data is interventional and generated from the DAG in Figure \ref{fig:figs2} c). For this model the error term is
\begin{equation*}
\delta_X(H,E,\epsilon_X) = \begin{cases} 
      f_X(H,\epsilon_X) - r(E) &\text{ if } E=-1\\
      x_1 - r(E)&\text{ if } E=1
     \end{cases},\\
\end{equation*}
and cannot be decomposed according to Assumption \ref{assump:decomp}. Inconsistency of the CD under do interventions was discussed in \cite{rothenhausler2019}. Since the Hybrid estimator also requires the error decomposition assumption, it will also be inconsistent for instruments/environments which are used to represent do operators. 

\subsection{Independent Instrumental Variables}
\label{sec:iiv}

Independent instrumental variable (IIV) models assume that $\delta_Y \ind E$. This can be leveraged to construct consistent estimates of $\beta_0$ even when $r(E) = 0$, much like the GCD \citep{saengkyongam2022exploiting,dunker2021adaptive,poirier2017efficient}. Note that $\delta_Y \ind E$ is equivalent to
\begin{equation}
\label{eq:iiv-moment}
\mathbb{E}[\eta(E)\phi(Y-X^T\beta_0)] = \mathbb{E}[\eta(E)]\mathbb{E}[\phi(Y-X^T\beta_0)]
\end{equation}
for all bounded, continuous functions $\eta$ and $\phi$. Thus estimates of $\beta_0$ may be constructed by finding $\beta$ which satisfy empirical versions of Equation \eqref{eq:iiv-moment} (up to sampling variability) for a large set of $\eta$ and $\phi$. See \cite{poirier2017efficient} for specific implementations of this approach.

Assumptions \ref{assump:causal-sem} do not require $\delta_Y \ind E$ because $\delta_Y(H,X,\epsilon_X)$ is a function of $X$ which is a function of $E$. One common setting where Assumptions \ref{assump:causal-sem} holds but $\delta_Y \notind E$ is heteroskedastic response models such as
\begin{equation*}
Y \gets X^T\beta_0 + H + s(X)\epsilon_Y
\end{equation*}
where $s : \mathbb{R}^p \rightarrow \mathbb{R}^+$ and $\mathbb{E}[\epsilon_Y] = 0$. Here IIV models should not be used.
}

%
%

\hypertarget{sim}{%
\section{Simulations}\label{sim}}

We demonstrate some of the applications of the GCD and Hybrid estimators in simulations.

\hypertarget{continuous-cd}{%
\subsection{Continuous Environments}\label{continuous-cd}}

We fit the GCD to a model with continuous environments/instruments. 
\begin{align*}
E &\gets Unif[0,1]\\
H &\gets N(0,1)\\
X &\gets 3H + (1 + 10E)\epsilon_X\\
Y &\gets X + 9H + \epsilon_Y
\end{align*}
where exogenous variables $\epsilon_X$ and $\epsilon_Y$ are standard normal. The true causal parameter is $\beta_0=1$. We simulate $n=100$ samples.

\begin{figure}
\centering
\includegraphics[width=0.5\linewidth]{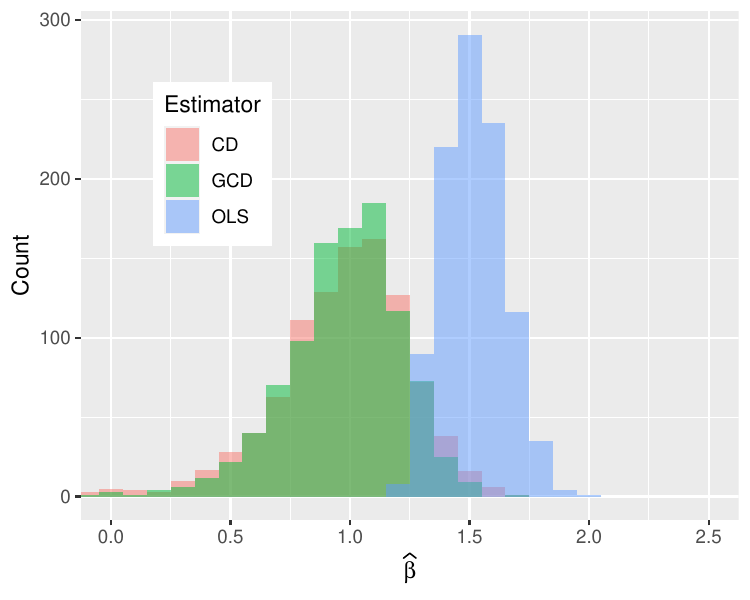}
\caption{The GCD sampling distribution is centered around the true causal effect with lower asymptotic variance than the CD. Ordinary Least Squares is inconsistent due to hidden confounding. IV (not shown) is also inconsistent because the instrument does not shift the mean of the exposures. \label{fig:cont-cd}}
\end{figure}

IV is not consistent for this model because $E$ shifts the variance of $X$ (larger $E$ implies larger variance for $X$), but not the mean of $X$. With univariate $E$ and $X$ the GCD is
\begin{equation}
\label{eq:cd-cont-1d}
\widehat{\beta}_{GCD} = \frac{\sum_{i=1}^n E_iX_iY_i}{\sum_{i=1}^nE_iX_i^2}.
\end{equation}
The CD is not directly applicable here because the environment is continuous. One could discretize $E$ with some function $e: \mathbb{R} \rightarrow \{0,1\}$ and then apply the CD using environment $e(E)$. We consider this approach with $e(E_i) = 1_{E_i > median(E_i)}$ (environment is 1 if $E_i$ is greater than sample median of $E$ values). We simulate $N=1000$ times and plot sampling distributions of the Generalized Causal Dantzig (GCD), the Causal Dantzig (CD), and Ordinary Least Squares (OLS) in Figure \ref{fig:cont-cd}. Note that $E$ is centered (mean shifted to $0$ in the sample) before the GCD is fit. OLS is inconsistent due to hidden confounding. The GCD and CD sampling distributions are both centered at the true causal effect of $1$. The GCD empirical sampling distribution is more concentrated around the causal effect. The GCD is also easier to fit because it does not require selection of the function $e$ to binarize the continuous variable $E$ into discrete environments.


\hypertarget{overid}{%
\subsection{Overidentified Models}\label{overid}}

The CD/GCD constraints overidentify $\beta_0$ whenever there is more than one instrument/environment. We consider data generating model:
\begin{align*}
E_1 &\gets Bernoulli(1/2)\\
E_2 &\gets Unif(0,1)\\
X_1 &\gets Y + X_2 + (1+3E_1)\epsilon_1\\
X_2 &\gets H + (1+3E_1+5E_2)\epsilon_2\\
X_3 &\gets H + X_1 + (1+5E_2)\epsilon_3\\
Y &\gets H + X_2 + \epsilon_Y.
\end{align*}
All exogenous variables $(H,\epsilon_1,\epsilon_2,\epsilon_3,\epsilon_Y)$ are standard normal. The true parameter value is $\beta_0=(0,1,0)$ because only $X_2$ has a causal effect on $Y$. The hidden confounder $H$ will cause OLS to be inconsistent. The sample size is $n=200$. We simulate $N=500$ runs. 

We consider four estimators: GCD (using both $E_1$ and $E_2$), GCDE1 (uses only environment $E_1$), GCDE2 (uses only environment $E_2$), and OLS (ordinary least squares). Note that GCDE1 is equivalent to the CD using $E_1$ to indicate the environment of the observation. GCDE2 is not equivalent to a CD estimator because $E_2$ is a continuous environment. For the GCD, the two--step estimator is used since the two environments overidentify $\beta_0$. For initial GCD weight matrix we use
\begin{equation*}
\widehat{W} = \frac{1}{n} (\bs{E \bullet X})^T (\bs{E \bullet X}).
\end{equation*}
Table \ref{tab:overid} shows the empirical coverage probabilities of 95\% confidence intervals for each parameter and the median CI width. The rows of the table correspond to different estimators. OLS is inconsistent because of hidden confounding. This results in the coverage probabilities being well below nominal levels ($0$ in the case of $\beta_2$). The three GCD methods (GCD, GCDE1, and GCDE2) all have empirical coverage probabilities near or above 95\%. However GCDE1 and GCDE2 obtain this coverage by producing extremely wide confidence intervals. For example, the median CI width for GCDE2 for $\beta_2$ is 10 times that for GCD. Similarly, the median width of the $\beta_3$ CI for GCDE1 is about 10 times the median width for GCD. By only using the information in one of the environment variables, GCDE1 and GCDE2 produce highly uncertain estimators with very wide confidence intervals.

\input{figs/gcd-overid}


\hypertarget{hybrid}{%
\subsection{Hybrid Estimator}\label{hybrid}}

\begin{figure}
   \centering
 a) \includegraphics[width=2.1in]{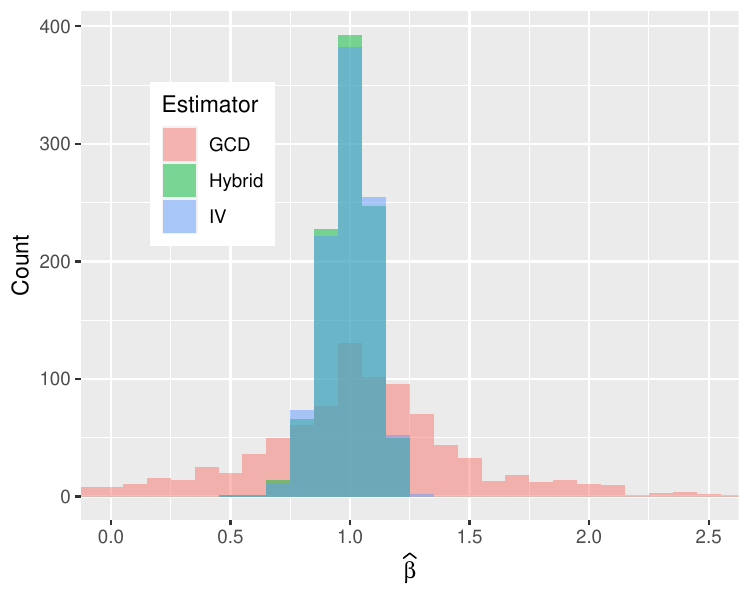}
 b) \includegraphics[width=2.1in]{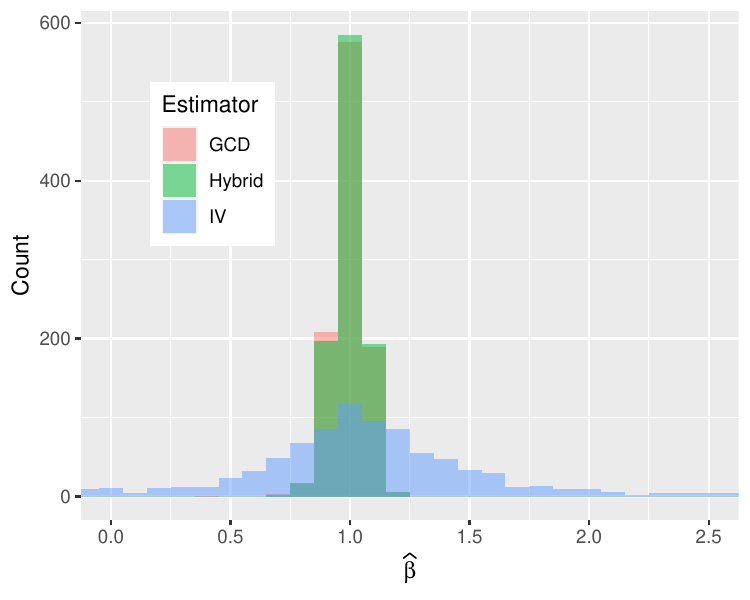}
\caption{a) \underline{Model 1:} IV and Hybrid dominate the GCD when the mean shift is strong but noise shift is weak. b) \underline{Model 2:} GCD and Hybrid dominate IV when the mean shift is weak but the noise shift is strong.\label{fig:sim_results2}}
 \end{figure}

Consider a model in which the instrument $E$ shifts the mean and variance of $X$:
\begin{align*}
E &\gets Unif[0,1]\\
H &\gets N(0,1)\\
X &\gets H + RE + (\alpha_v E + \alpha_0)\epsilon_X\\
Y &\gets H + \beta_0 X + \epsilon_Y.
\end{align*}
IV is inconsistent when $R=0$. It will be a poor estimator when $R$ is near $0$ because the instrument is weak. In similar fashion GCD is inconsistent when $\alpha_v = 0$. It will be a poor estimator when $\alpha_v$ is near $0$. The Hybrid estimator is consistent whenever IV or GCD is consistent because it can leverage changes in the mean or variance induced by the instrument/environment $E$. Here we consider two simulation settings: In \underline{Model 1}, $R=5$ and $\alpha_v=1$ so that $E$ has a strong effect on the mean of $X$ and only a weak effect on the variance. In \underline{Model 2}, $R=1$ and $\alpha_v=5$ so that $E$ has a weak effect on the mean of $X$ and a strong effect on the variance. We fit IV, GCD, and the Hybrid estimator on these two models. IV and GCD do not require specification of a weight matrix because the number of parameters equals the number of constraints. For the Hybrid estimator, we perform a two step procedure to estimate the optimal weight matrix, using 
\begin{equation*}
\widehat{W} = \frac{1}{n} (\bs{E},\bs{E}\bullet\bs{X})^T (\bs{E},\bs{E}\bullet\bs{X})
\end{equation*}
for an initial weighting. We let $\beta_0=1$ and simulate $n=200$ samples. Empirical sampling distributions with $N=1000$ simulations are shown in Figure \ref{fig:sim_results2}. The Hybrid estimator performs well for both models while IV and the GCD each only perform well for one of the models.


\hypertarget{application}{%
\section{Application to Flow Cytometry Data}\label{application}}

\cite{sachs2005causal} measured the abundance of 11 biochemical agents in thousands of cells using flow cytometry. These data were collected under several conditions (or environments) in which external reagents were added to the system. Each reagent has the effect of stimulating or inhibiting particular agents in the system. Five conditions used in this work (1 observational and 4 interventional) are described in Table \ref{tab: data-description}. In the observational condition, only a general perturbation (CD3+CD28) is applied. For the observational condition, the expression of the 11 biochemical agents was measured in 853 cells.  In Condition 3, the reagent Psitectorignin, an inhibitor of PIP2 was added to the system in addition to the general perturbation. The effect of this perturbation should reduce the abundance of PIP2 (one of the 11 measured agents) as well as alter the abundance of any agents which PIP2 itself effects. 

\input{figs/data-description}

\cite{meinshausen2016methods} fit the CD to this data to infer a causal signalling network (graph) among the 11 agents. Here we compare the performance of the Causal Dantzig with IV, GCD, and Hybrid estimators. Before fitting any models, we hyperbolic arcsine transform the data. This technique is used to approximately normalize the flow cytometry data to better satisfy modelling assumptions and reduce the influence of outliers \citep{ray2012transformation}.

\subsection{Univariate Analysis}

We first consider the problem of determining whether a particular agent $X$ has a total causal effect on another agent $Y$. In general, simple regression of $Y$ on $X$ will not consistently estimate the total causal effect of $X$ on $Y$ because hidden confounding and reverse causality will induce an association between $X$ and $Y$ even when $X$ has no causal effect on $Y$. To address this problem, we consider $(X,Y)$ data from two environments: an observational environment involving only a general system perturbation and an interventional environment which includes the general perturbation plus a reagent designed to perturb $X$. Presence or absence of the additional reagent is modeled using an instrument (environment) variable $E$. We focus on two causes (different $X$ variables), PIP2 and MEK, in order to demonstrate similarities and differences in IV and CD modeling results.


\subsubsection{PIP2}




Using the Causal Dantzig, \cite{meinshausen2016methods} (Figure 3) did not find that PIP2 is a direct cause of changes in any of the other 10 biochemical agents. This implies that PIP2 should not have a total effect on any of the agents in the system. To investigate this, we consider abundance measures from cells in two conditions: observational (the general perturbation only) and Condition 3 (Psitectorignin plus general perturbation). The condition is treated as a binary instrumental variable / environment. This is justified by the fact that Psitectorignin is meant to directly target PIP2 and any effects on other agents should occur by way of PIP2. Since the CD and GCD are identical in this case, we compare the CD, IV, and Hybrid estimators.

\begin{figure}
   \centering
 a) \includegraphics[width=2.1in]{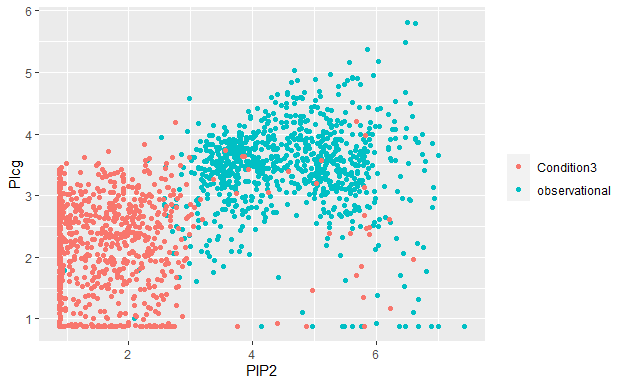}
 b) \includegraphics[width=2.1in]{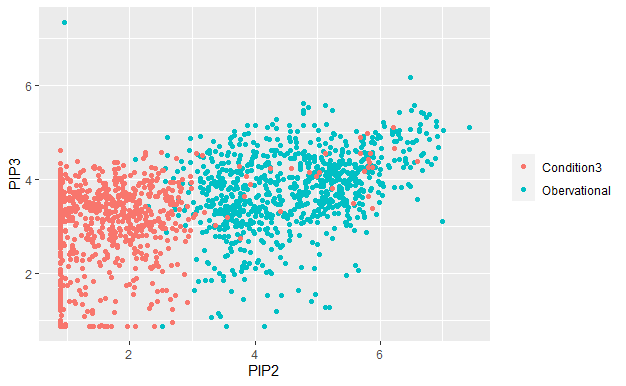}
\caption{a) Scatter plot for PIP2 vs Plcg b) Scatter plot for PIP2 vs PIP3.\label{fig:exp3}}
 \end{figure}

Figure \ref{fig:exp3} displays scatter plots of cellular abundances of Plcg versus PIP2 and PIP3 versus PIP2. Red points are for cells measured with the intervention Psitectorignin applied while blue points are cells measured without the intervention.  As expected, Psitectorignin has a strong effect on PIP2, substantially decreasing its mean. Plcg and PIP3 abundances are also strongly influenced by the intervention. This suggests that PIP2 is a cause (either directly or possibly indirectly through other agents) of both Plcg and PIP3. Note that the intervention primarily effects the mean, rather than the variance, of PIP2. Thus the CD may struggle to identify an effect because it is sensitive to variance, not mean, shifts. In contrast, IV is better suited to settings where the instrument/environment effects the exposure mean. This is a possible explanation for why the CD did not identify PIP2 as a cause of changes of other agents in \cite{meinshausen2016methods}.

Table \ref{tab: result-pip2} contains parameter estimates, confidence intervals, and p--values for the CD, IV and Hybrid estimators fit to the data in Figure \ref{fig:exp3}. As expected, the CD does not find a significant causal effect while IV does. The Hybrid estimator, which can leverage changes in mean or variance, produces estimates very similar to the IV.

\input{figs/result-pip2}

\subsubsection{MEK}

We now consider estimating the total causal effect of MEK on RAF, using data from the observational condition and condition 4. Since the condition 4 reagent targets MEK, this serves as a good instrument. Using the Causal Dantzig, \cite{meinshausen2016methods} found that MEK has a direct effect on RAF.

\begin{figure}
   \centering
 \includegraphics[width=2.7in]{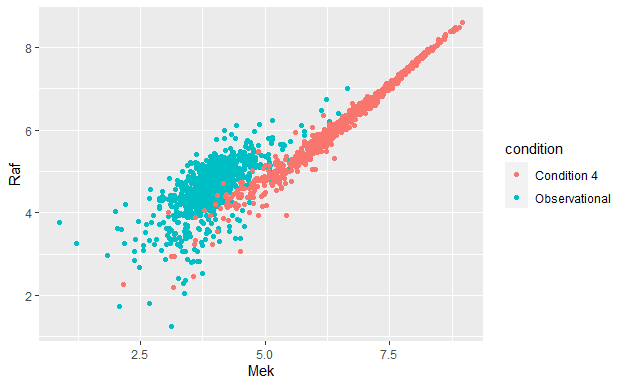}
\caption{Scatter plot that shows the relation between Mek and Raf. \label{fig:mek-raf}}
 \end{figure}

Figure \ref{fig:mek-raf} shows the scatter plot of MEK versus RAF. The mean of MEK in condition 4 is higher than in the observational condition. Further the variance of MEK has increased in condition 4 relative to the observational condition. Thus both CD and IV are likely suitable to estimating a causal effect in this situation. Table \ref{tab: mek-raf} shows IV, CD and Hybrid parameter estimates, confidence intervals, and p--values fit using the data in Figure \ref{fig:mek-raf}. All three estimators identify a causal effect.

\input{figs/exp1-res}

\subsection{Multivariate Analysis} 

For the multivariate analysis, we fit models using the 5 conditions specified in Table \ref{tab: data-description}. \revision{Each condition is treated as a data generating environment. Thus there are 5 environments. An instrument/environment variable $E \in \mathbb{R}^4$ is created following the procedure outlined in Equation \eqref{eq:iv-env-const}.} We iteratively treat each of the 11 agents as a response and regress it on the other 10 agent abundance measurements. Due to the limited number of conditions relative to reagents (4 instruments and 10 exposures), the IV can not be directly used in this situation. However, we would like the estimator to be sensitive to shifts in mean induced by the interventions. Thus we fit a Hybrid estimator specified in Equation \eqref{eq:hybrid}. We compare the performance of the Hybrid estimator with the CD and GCD.

Note that, if an agent is used as the target or response variable, then intervention on that agent is not allowed in either CD, GCD or Hybrid. \revision{This is equivalent to an instrument having a direct effect on the response and would violate Assumption \ref{assump:consist}b.} So when one of Akt, PIP2, Mek or PKC is used as the response, the total number of environments is 4 (observational and 3 conditions), while for all the other cases, the number of environments is 5 (observational and 4 conditions). \revision{For constructing the weight matrices, we use two-step estimators proposed Theorem \ref{thm:cd-gmm-est-an} with initial weight matrices as described in Section \ref{overid} for the GCD and Section \ref{hybrid} for the Hybrid.}

When the number of environments is greater than two, the GCD and the CD are not equivalent. \cite{rothenhausler2019} proposed two methods for handling the case with greater than 2 environments. We use the \texttt{hiddenICP} version of the CD from the R package \texttt{InvariantCausalPrediction}.  With $K>2$ environments, \texttt{hiddenICP} iteratively fits the CD with one environment versus all the other environments ($K$ fits). This produces $K$ point estimates and $K$ confidence intervals for each parameter. The parameter estimates across the $K$ fits are averaged to create a single point estimate. The confidence intervals lower limit is the smallest of the lower limits of the individual CD fit confidence intervals. Likewise the confidence interval upper limit is the largest of the upper limits of the individual confidence intervals. Thus the intervals are conservative.



\begin{figure}
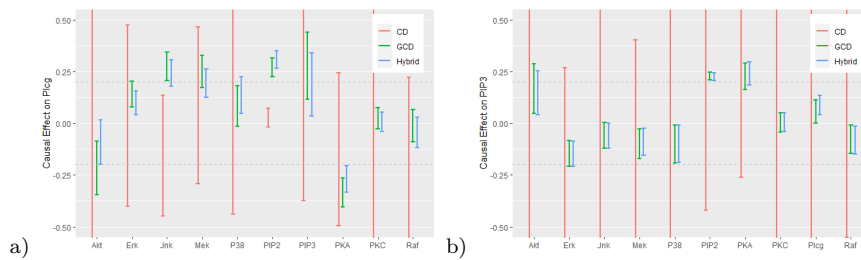

   \centering
 a) \includegraphics[width=2.1in]{figs/plcg.pdf}
 b) \includegraphics[width=2.1in]{figs/pip3.pdf}
\caption{a) Hybrid, GCD and CD estimation results when Plcg is the response.  b) Hybrid, GCD and CD estimation results when PIP3 is the response.  \label{fig:plcg-pip3}}
 \end{figure}

Figure \ref{fig:plcg-pip3} compares 95\% confidence intervals for the CD, GCD, and Hybrid estimators with a) Plcg and b) PIP3 as the response. CD confidence intervals are very wide relative to GCD and Hybrid. The GCD and the Hybrid estimators perform similarly. In our univariate analysis, we found that PIP2 had a total effect on both Plcg and PIP3. In the multivariate analysis here, GCD and Hybrid identify PIP2 as direct causes of changes in both Plcg and PIP3. The CD fails to identify this effect, due to suboptimal merging of environments and an excessively conservative strategy for constructing confidence intervals.

\begin{figure}
   \centering
      \includegraphics[width=3.1in]{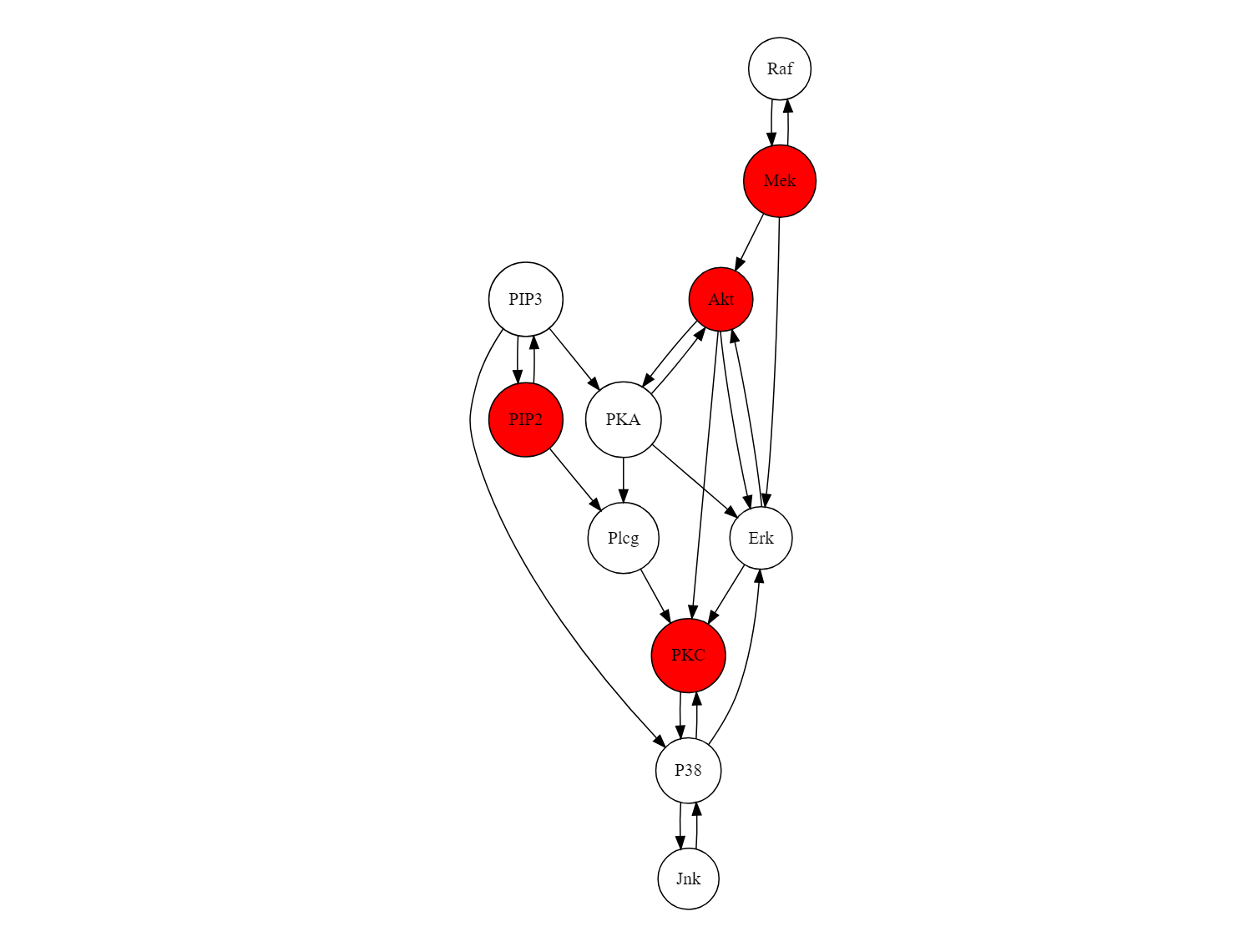}
\caption{Causal network found by the Hybrid estimator. The red circles represent the agents on which interventions were applied. \label{fig:multi}}
 \end{figure}

We define a causal effect as strong if the entire 95\% confidence interval is outside the range of $(-.2,.2)$. Figure \ref{fig:multi} shows the strong causal relations found by the Hybrid estimator in a graph. Compared with the results from \cite{meinshausen2016methods} (Figure 3), our hybrid estimator finds more strong causal relations: 24 strong relations for hybrid versus 13 for CD. Some relations are found by both methods, e.g. the causal effects and the reverse causal effects between Raf--Mek, P38--Jnk, and Erk--Akt. \revision{The Hybrid model identifies the entire Raf->Mek->Erk path which was seen as a major validation of the original Bayesian network applied to the system \citep{sachs2005causal} while the original application of the CD failed to identify the Mek->Erk edge.} Further, the CD did not discover any causal effects from PIP2 to Plcg and PIP2 to PIP3, which are illustrated to exist from Figure \ref{fig:exp3} and discovered by our hybrid estimator. In addition, the hybrid estimator also found similar causal relations with other methods (ICP from  \cite{peters2016causal} and the Bayesian network from \cite{sachs2005causal}). For example, ICP and the hybrid both discovered the causal relations from PIP2 to Plcg and PKA to Erk, while the Bayesian network and the hybrid both discovered the relations from Plcg to PKC and PIP2 to PKC.

\hypertarget{discussion}{%
\section{Discussion}\label{discussion}}

\revision{
In this work, we proposed two new methods, the Generalized Causal Dantzig (GCD) and Hybrid, for estimating causal effects in the presence of hidden confounding and reverse causality. The GCD generalizes the Causal Dantzig estimator of \cite{rothenhausler2019} to problems with continuous environments. Further we developed theory (based on GMM estimation) for the GCD in the over--identified case. This was not present in the original CD paper. The Hybrid estimator enforces both GCD environment and instrumental variable (IV) moment constraints, further illustrating connections between the concepts of environment and instrument which have been previously noted \citep{heinze2018invariant,peters2016causal}. We demonstrated the utility of these estimators in simulations and an application to Flow Cytometry data.

In this work, we did not discuss high--dimensional estimation. \cite{rothenhausler2019} proposed using the Dantzig selector \citep{candes2007dantzig} to regularize the Causal Dantzig in high dimensional problems. This required extensive theoretical development. Since both the GCD and Hybrid are are GMM estimators, existing methods for fitting high dimensional GMM estimators (penalty terms, tuning parameter selection algorithms, and theory) may be applicable for the GCD and Hybrid estimators \citep{belloni2018high}. This represents a direction for future research. The performance of these high--dimensional estimators may be compared with existing high--dimensional instrumental variable estimators \citep{belloni2012sparse,gold2020inference,lin2015regularization}.
}

\hypertarget{proofs}{%
\section{Supplementary Material}\label{proofs}}

\revision{
\subsection{CD and Lewbel \cite{lewbel2012using}\label{lewbel}}

We demonstrate equivalence between Lewbel's GMM (Equation 12 of \cite{lewbel2012using}) and the CD/GCD for the case of univariate $X$ and some restrictions on Lewbel's model (Equations 1 and 2 of \cite{lewbel2012using}). With univariate $X$, the GMM defining the GCD (Equation \eqref{eq:cd-gmm-const}) becomes
\begin{equation}
\label{eq:gcd-lewbel}
m_{GCD}(\beta) =  \mathbb{E}[EX(Y-X^T\beta)].
\end{equation}
Translate from Lewbel notation to our notation according to:
\begin{align*}
&0 \gets X\\
&E \gets Z\\
&X \gets Y_2\\
&\beta \gets \gamma_1.
\end{align*}
Lewbel's $X$ is a vector of observed exogenous regressors which we do not consider (thus $0$). Assume Lewbel's $\gamma_2=0$ (triangular system) and $\mu=\mathbb{E}[Z]=0$ (environment/instrument $E$ is mean $0$). Under these restrictions and notational changes, $Q_1$, $Q_2$ and $Q_3$ of Lewbel Section 3.1 are $0$ and $Q_4=EX(Y-X^T\beta)$. Thus Equation \eqref{eq:gcd-lewbel} equals Equation 12 of Lewbel. Hence the estimators are equivalent.

}

\hypertarget{cd-gmm-proof}{%
\subsection{Proof of Theorem \ref{thm:cd-gmm}}\label{cd-gmm-proof}}

Recall $q = \#\mathcal{E}-1$. We show that for any $k \in \{1,\ldots,p\}$
\begin{align*}
&\cap_{e=1}^q \{ \beta : m_{GCD}(\beta)_{e+q(k-1)}=0\}\\ 
&= \{ \beta : \mathbb{E}[X_k^f(Y^f - X^{f^T}\beta)] = \mathbb{E}[X_k^g(Y^g - X^{g^T}\beta)] \, \forall \, f,g\}.
\end{align*}
This implies the result because
\begin{align*}
&\{ \beta : m_{GCD}(\beta) = 0\} \\
=& \cap_{k=1}^{p} \cap_{e=1}^q \{ \beta : m_{GCD}(\beta)_{e+q(k-1)}=0\} \\
=&  \cap_{k=1}^{p}  \{ \beta : \mathbb{E}[X_k^f(Y^f - X^{f^T}\beta)] = \mathbb{E}[X_k^g(Y^g - X^{g^T}\beta)] \, \forall \, f,g\}\\
=&  \{ \beta : \mathbb{E}[X^f(Y^f - X^{f^T}\beta)] = \mathbb{E}[X^g(Y^g - X^{g^T}\beta)] \, \forall \, f,g\}.
\end{align*}
Note that
\begin{align*}
m_{GCD}(\beta)_{e+q(k-1)} &= \mathbb{E}[E_eX_k(Y-X^T\beta)]\\
&= \sum_{l=0}^1 \mathbb{E}[s_{el}X_k(Y-X^T\beta)|E_e=s_{el}]P(E_e=s_{el}).
\end{align*}
Noting that Equation \eqref{eq:iv-env-const} implies $(s_{e1}P(E_e=s_{e1})) / (s_{e0}P(E_e=s_{e0})) = -1$, we have 
\begin{align*}
&\{\beta : m_{GCD}(\beta)_{e+q(k-1)}=0\} \\
=& \{\beta : \mathbb{E}[X_k(Y-X^T\beta)|E_e=s_{e1}] = \mathbb{E}[X_k(Y-X^T\beta)|E_e=s_{e0}]\}\\
=& \{\beta : \mathbb{E}[X_k(Y-X^T\beta)|\mathcal{E}=e] = \mathbb{E}[X_k(Y-X^T\beta)|\mathcal{E}\neq e]\}\\
=& \{\beta : \mathbb{E}[X_k(Y-X^T\beta)|\mathcal{E}=e] = \sum_f \mathbb{E}[X_k(Y-X^T\beta)|\mathcal{E} = f]P(\mathcal{E}=f|\mathcal{E}\neq e) \}.
\end{align*}
Let $b \in \mathbb{R}^{q}$ with $b_e \equiv \mathbb{E}[X_k(Y-X^T\beta)|\mathcal{E}=e]$, $A \in \mathbb{R}^{q \times (q+1)}$ with $A_{e,f} = P(\mathcal{E}=f|\mathcal{E}\neq e)$, and $c \in \mathbb{R}^1$ with $c=\mathbb{E}[X_k(Y-X^T\beta)|\mathcal{E}=\#\mathcal{E}]$. Then
\begin{equation*}
\cap_{e=1}^q \{\beta : m_{GCD}(\beta)_{e+q(k-1)}=0\} = \left \{\beta : b = A\begin{pmatrix}b\\c\end{pmatrix}\right\}.
\end{equation*}
So sufficient to show
\begin{equation*}
\left \{\beta : b = A\begin{pmatrix}b\\c\end{pmatrix}\right\} = \{ \beta : \mathbb{E}[X_k^f(Y^f - X^{f^T}\beta)] = \mathbb{E}[X_k^g(Y^g - X^{g^T}\beta)] \, \forall \, f,g\}.
\end{equation*}
The r.h.s. implies that $b_e=b_f$ for all $e,f$ and $c=b_e$ for all $e$. Since the rows of $A$ sum to $1$, this implies the l.h.s. set condition is satisfied. Thus the r.h.s. is a subset of the l.h.s. Now we show set equivalence by arguing that if $\beta$ is not in the r.h.s., it is not in the l.h.s. Note that all elements of $A$ are non--negative and the rows of $A$ sum to $1$. If $\beta$ is not in the r.h.s. set then either:

\begin{enumerate}
\item $\max b > c$. Select $m \in \argmax{e \in \{1,\ldots,q\}} b_e$. Then
\begin{align*}
b_m &= \sum_{e=1}^q A_{me}b_m + A_{m,q+1}b_m\\
&> \sum_{e=1}^q A_{me}b_e + A_{m,q+1}c.
\end{align*}
Thus the l.h.s. equality constraints are not satisfied.
\item $\min b < c$. Select $m \in \argmin{e \in \{1,\ldots,q\}} b_e$. Then
\begin{align*}
b_m &= \sum_{e=1}^q A_{me}b_m + A_{m,q+1}b_m\\
&< \sum_{e=1}^q A_{me}b_e + A_{m,q+1}c.
\end{align*}
Thus the l.h.s. equality constraints are not satisfied.
\end{enumerate}

\hypertarget{cd-gmm-est-proof}{%
\subsection{Proof of Theorem \ref{thm:cd-gmm-est}}\label{cd-gmm-est-proof}}

\begin{enumerate}
\item Note that
\begin{align*}
&||\widehat{m}_{GCD}(\beta)||^2_{\widehat{W}} \\
=& \beta^T\overbrace{\bs{X}^T(\bs{E} \bullet \bs{X}) \widehat{W} (\bs{E} \bullet \bs{X})^T \bs{X}}^{\equiv M} \beta - 2 \beta^T\bs{X}^T (\bs{E} \bullet \bs{X})\widehat{W}(\bs{E} \bullet \bs{X})^T\bs{Y} + C
\end{align*}
where $C$ does not depend on $\beta$. This is a quadratic function in $\beta$. The matrix $M$ is positive definite under the assumptions that $\widehat{W} \succ 0$ and $\bs{X}^T(\bs{E} \bullet \bs{X})$ has column rank $p$. Thus the unique minimizer is given by 
\begin{equation*}
\widehat{\beta}_{GCD}(\widehat{W}) = (\bs{X}^T(\bs{E}\bullet\bs{X})\widehat{W}(\bs{E}\bullet\bs{X})^T\bs{X})^{-1}(\bs{X}^T(\bs{E}\bullet \bs{X})\widehat{W}(\bs{E} \bullet \bs{X})^T \bs{Y}).
\end{equation*}
\item Define $\bs{P} = \bs{X}^T(\bs{E}\bullet\bs{X})\widehat{W}^{1/2} \in \mathbb{R}^{p \times pr}$. When $q=1$, $\bs{P}$ is invertible, so we have
\begin{align*}
\widehat{\beta}_{GCD}(\widehat{W}) &= (\bs{P}\bs{P}^T)^{-1}(\bs{P}W^{1/2}(\bs{E} \bullet \bs{X})^T \bs{Y})\\
&= (\bs{P}^T)^{-1}(W^{1/2}(\bs{E} \bullet \bs{X})^T \bs{Y})\\
&= ((\bs{E} \bullet \bs{X})^T \bs{X})^{-1} (\bs{E} \bullet \bs{X})^T\bs{Y}.
\end{align*}
\item Recall that $\sum E_i = 0$ which implies that $n_1 s_{11} + n_2 s_{10}=0$. Then we have
\begin{align*}
\widehat{\beta}_{GCD} =& ((\bs{E} \bullet \bs{X})^T \bs{X})^{-1} (\bs{E} \bullet \bs{X})^T\bs{Y}\\
=& \left(\frac{1}{n} \sum E_iX_iX_i^T\right)^{-1}\left(\frac{1}{n}\sum E_i X_iY_i\right)\\
=& \left(\frac{s_{11}}{n} \sum X_iX_i^T 1_{E_i=s_{11}} + \frac{s_{10}}{n} \sum X_iX_i^T 1_{E_i=s_{10}}\right)^{-1}\\
&\times\left(\frac{s_{11}}{n}\sum X_iY_i1_{E_i=s_{11}} + \frac{s_{10}}{n}\sum X_iY_i1_{E_i=s_{10}}\right)\\
=& \left(s_{11} \sum X_iX_i^T 1_{E_i=s_{11}} - \frac{n_1 s_{11}}{n_2} \sum X_iX_i^T 1_{E_i=s_{10}}\right)^{-1}\\
&\times \left(s_{11}\sum X_iY_i1_{E_i=s_{11}} - \frac{n_1 s_{11}}{n_2} \sum X_iY_i1_{E_i=s_{10}}\right)\\
=& \left(\frac{1}{n_1}\bs{X}^{1^T}\bs{X}^1 - \frac{1}{n_2} \bs{X}^{2^T}\bs{X}^2\right)^{-1} \left(\frac{1}{n_1}\bs{X}^{1^T}\bs{Y}^1 - \frac{1}{n_2}\bs{X}^{2^T}\bs{Y}^2\right)\\
=& \widehat{\beta}_{CD}.
\end{align*}
\end{enumerate}

\hypertarget{cd-gmm-est-an-proof}{%
\subsection{Proof of Theorem \ref{thm:cd-gmm-est-an}}\label{cd-gmm-est-an-proof}}

Sufficient to show that
\begin{equation*}
\widehat{W}_{GCD} \rightarrow_P \bar{W}.
\end{equation*}
Let $\gamma_i$ be the $i^{th}$ row of $\bs{E \bullet X}$. Then
\begin{align*}
\widehat{W}_{GCD}^{-1} =& \frac{1}{n} \sum_{i=1}^n \gamma_{i}\gamma_{i}^T\widehat{\bs{\Sigma}}_{ii}\\
=& \frac{1}{n} \sum_{i=1}^n \gamma_{i}\gamma_{i}^T(Y_i - X_i^T\beta_0 + X_i^T(\beta_0-\widehat{\beta}_{GCD}(\widehat{W})))^2\\
=& \underbrace{\frac{1}{n} \sum_{i=1}^n \gamma_{i}\gamma_{i}^T(Y_i - X_i^T\beta_0)^2}_{\equiv B_1} \\
+& \underbrace{\frac{2}{n} \sum_{i=1}^n \gamma_{i}\gamma_{i}^T(Y_i - X_i^T\beta_0)X_i^T(\beta_0-\widehat{\beta}_{GCD}(\widehat{W}))}_{\equiv B_2} \\
+& \underbrace{\frac{1}{n} \sum_{i=1}^n \gamma_{i}\gamma_{i}^T(X_i^T(\beta_0-\widehat{\beta}_{GCD}(\widehat{W})))^2}_{\equiv B_3}.
\end{align*}
The quantity $B_1 \rightarrow_P \bar{W}^{-1}$ by the LLN. Sufficient to show $B_2,B_3 \rightarrow_P 0$. Note that
\begin{equation*}
||B_2||_\infty \leq 2 ||\beta_0-\widehat{\beta}_{GCD}(\widehat{W})||_\infty\left(\frac{1}{n}\sum_{i=1}^n ||\gamma_{i}\gamma_{i}^T(Y_i - X_i^T\beta_0)||_\infty ||X_i||_\infty\right) \rightarrow_P 0.
\end{equation*}
Similarly
\begin{equation*}
||B_3||_\infty \leq ||\beta_0-\widehat{\beta}_{GCD}(\widehat{W})||^2_\infty\left(\frac{1}{n}\sum_{i=1}^n ||\gamma_{i}\gamma_{i}^T||_\infty ||X_i||_\infty^2\right) \rightarrow_P 0.
\end{equation*}

\hypertarget{consist}{%
\subsection{Proof of Theorem \ref{thm:consist}}\label{consist}}

Note that for all $x$ we have
\begin{align*}
0 &= \mathbb{E}[\delta_Y(H,X,\epsilon_Y)|do(X=x)]\\
&= \mathbb{E}[\delta_Y^1(H,\epsilon_Y)|do(X=x)] + \mathbb{E}[\delta_Y^2(X,\epsilon_Y)|do(X=x)]\\
&= \mathbb{E}[\delta_Y^1(H,\epsilon_Y)] + \mathbb{E}[\delta_Y^2(x,\epsilon_Y)]
\end{align*}
where the last equality holds because there are no backdoor paths from $X$ to $(H,\epsilon_Y)$ or from $X$ to $(X,\epsilon_Y)$. Thus there is constant $c$ such that $c=\mathbb{E}[\delta_Y^2(x,\epsilon_Y)]$ for all $x$ and $-c=\mathbb{E}[\delta_Y^1(H,\epsilon_Y)]$.

Next note that $Z_1 \ind \epsilon_Y | Z_2$ for all sets of variables $Z_1$ and $Z_2$ which do not contain $Y$ because all paths from $Z_1$ to $\epsilon_Y$ must contain collider $Y$ which is not in $Z_2$ (by assumption) and has no descendants (\cite{pearl2009causal} d--Separation Definition 1.2.3). Thus we have

\begin{align*}
E &\ind \epsilon_Y | X\\
E,H &\ind \epsilon_Y | X,\epsilon_X\\
\epsilon_X &\ind \epsilon_Y | X
\end{align*}

\begin{itemize}
\item For the IV estimator
\begin{align*}
m_{IV}(\beta) &= \mathbb{E}[E(Y-X^T\beta)]\\
&= \mathbb{E}[EX^T](\beta_0-\beta) + \mathbb{E}[E\delta_Y(H,X,\epsilon_Y)]
\end{align*}
Sufficient to show that $\mathbb{E}[E\delta_Y(H,X,\epsilon_Y)]=0$. We have
\begin{align*}
\mathbb{E}[E\delta_Y(H,X,\epsilon_Y)] &= \mathbb{E}[E\delta_Y^1(H,\epsilon_Y)] + \mathbb{E}[E\delta_Y^2(X,\epsilon_Y)]\\
&= \underbrace{\mathbb{E}[E]}_{=0}\mathbb{E}[\delta_Y^1(H,\epsilon_Y)] + \mathbb{E}[E\delta_Y^2(X,\epsilon_Y)]\\
&= \mathbb{E}[\mathbb{E}[E|X]\mathbb{E}[\delta_Y^2(X,\epsilon_Y)|X]]\\
&= \mathbb{E}[\mathbb{E}[E|X]\mathbb{E}[\delta_Y^2(X,\epsilon_Y)|do(X)]]\\
&= \mathbb{E}[\mathbb{E}[E|X]c]\\
&= 0
\end{align*}
\item For the GCD estimator
\begin{align*}
m_{GCD}(\beta) &= \mathbb{E}[vec(EX^T)(Y-X^T\beta)]\\
&= \mathbb{E}[vec(EX^T)X^T](\beta_0-\beta) + \mathbb{E}[vec(EX^T)\delta_Y(H,X,\epsilon_X)]
\end{align*}
Sufficient to show that $\mathbb{E}[vec(EX^T)\delta_Y(H,X,\epsilon_X)]=0$. Let $i \in \{1,\ldots,q\}$ and $j \in \{1,\ldots,p\}$. Elements of the vector are of the form
\begin{align*}
&\mathbb{E}[E_iX_j\delta_Y(H,X,\epsilon_Y)] \\
=& \mathbb{E}[E_i(r(E)_j + \delta_X^1(E,\epsilon_X)_j + \delta_X^2(H,\epsilon_X)_j)(\delta_Y^1(H,\epsilon_Y) + \delta_Y^2(X,\epsilon_Y)].
\end{align*}
First note
\begin{align*}
&\mathbb{E}[E_ir(E)_j(\delta_Y^1(H,\epsilon_Y) + \delta_Y^2(X,\epsilon_Y))] \\
=& \mathbb{E}[E_ir(E)_j\delta_Y^1(H,\epsilon_Y)]  + \mathbb{E}[E_ir(E)_j\delta_Y^2(X,\epsilon_Y))]\\
=& - c\mathbb{E}[E_ir(E)_j] + \mathbb{E}[\mathbb{E}[E_ir(E)_j|X]\mathbb{E}[\delta_Y^2(X,\epsilon_Y))|X]]\\
=& - c\mathbb{E}[E_ir(E)_j] + \mathbb{E}[\mathbb{E}[E_ir(E)_j|X]\mathbb{E}[\delta_Y^2(X,\epsilon_Y))|do(X)]]\\
=& - c\mathbb{E}[E_ir(E)_j] + c\mathbb{E}[E_ir(E)_j]\\
=& 0.
\end{align*}
Next we have
\begin{align*}
& \mathbb{E}[E_i\delta_X^1(E,\epsilon_X)_j(\delta_Y^1(H,\epsilon_Y) + \delta_Y^2(X,\epsilon_Y))] \\
=& \mathbb{E}[E_i\delta_X^1(E,\epsilon_X)_j\delta_Y^1(H,\epsilon_Y)]  + \mathbb{E}[E_i\delta_X^1(E,\epsilon_X)_j\delta_Y^2(X,\epsilon_Y))]\\
=& - c\mathbb{E}[E_i\delta_X^1(E,\epsilon_X)_j] + \mathbb{E}[\mathbb{E}[E_i\delta_X^1(E,\epsilon_X)_j|X,\epsilon_X]\mathbb{E}[\delta_Y^2(X,\epsilon_Y))|X,\epsilon_X]]\\
=& - c\mathbb{E}[E_i\delta_X^1(E,\epsilon_X)_j] + \mathbb{E}[\mathbb{E}[E_i\delta_X^1(E,\epsilon_X)_j|X,\epsilon_X]\mathbb{E}[\delta_Y^2(X,\epsilon_Y))|do(X)]]\\
=& - c\mathbb{E}[E_i\delta_X^1(E,\epsilon_X)_j] + \mathbb{E}[\mathbb{E}[E_i\delta_X^1(E,\epsilon_X)_j|X,\epsilon_X]c]\\
=& - c\mathbb{E}[E_i\delta_X^1(E,\epsilon_X)_j] + c\mathbb{E}[E_i\delta_X^1(E,\epsilon_X)_j]\\
=& 0.
\end{align*}
Finally
\begin{align*}
&\mathbb{E}[E_i\delta_X^2(H,\epsilon_X)_j(\delta_Y^1(H,\epsilon_Y) + \delta_Y^2(X,\epsilon_Y))] \\
=& \mathbb{E}[E_i\delta_X^2(H,\epsilon_X)_j\delta_Y^1(H,\epsilon_Y)]\\
+&\mathbb{E}[E_i\delta_X^2(H,\epsilon_X)_j\delta_Y^2(X,\epsilon_Y))]\\
=& \mathbb{E}[E_i]\mathbb{E}[\delta_X^2(H,\epsilon_X)_j\delta_Y^1(H,\epsilon_Y)]\\
+& \mathbb{E}[\mathbb{E}[E_i\delta_X^2(H,\epsilon_X)_j|X,\epsilon_X]\mathbb{E}[\delta_Y^2(X,\epsilon_Y))|X,\epsilon_X]]\\
=& c\mathbb{E}[E_i\delta_X^2(H,\epsilon_X)_j]\\
=& 0.
\end{align*}

\item For the Hybrid estimator
\begin{align*}
m_{H}(\beta) &= \mathbb{E}\left[\begin{pmatrix} E \\ vec(EX^T) \end{pmatrix}(Y-X^T\beta)\right]\\
&= \mathbb{E}\left[\begin{pmatrix} E \\ vec(EX^T) \end{pmatrix}X^T\right](\beta_0-\beta) + \mathbb{E}\left[\begin{pmatrix} E \\ vec(EX^T) \end{pmatrix}\delta_Y\right].
\end{align*}
Sufficient to show $\mathbb{E}\left[\begin{pmatrix} E \\ vec(EX^T) \end{pmatrix}\delta_Y\right]=0$. Each element of this vector was shown to be mean $0$ in the proofs for $m_{IV}$ and $m_{GCD}$.
\end{itemize}

\hypertarget{ack}{%
\section*{Acknowledgements}\label{ack}}

James P. Long was partially supported by National Institutes of Health SPORE [P50CA127001, P50CA140388] and CCTS [UL1TR003167]. Kim-Anh Do was partially supported by the National Institutes of Health [P30CA016672], SPORE [P50CA140388], CCTS [TR000371] and CPRIT [RP160693].

\bibliographystyle{abbrvnat}
\bibliography{refs}

\end{document}

%% file: figs/gcd-overid.tex
\begin{table}

\caption{Coverage and Median Width of Confidence Intervals for Different Estimators \label{tab:overid}}
\centering
\begin{tabular}[t]{l|r|r|r|r|r|r}
\hline
\multicolumn{1}{c|}{ } & \multicolumn{3}{c|}{Coverage} & \multicolumn{3}{c}{Median Width} \\
\cline{2-4} \cline{5-7}
  & $\beta_1$ & $\beta_2$ & $\beta_3$ & $\beta_1$ & $\beta_2$ & $\beta_3$\\
\hline
GCD & 0.94 & 0.96 & 0.94 & 0.25 & 0.39 & 0.16\\
\hline
GCDE1 & 1.00 & 0.99 & 1.00 & 1.61 & 0.63 & 1.68\\
\hline
GCDE2 & 0.98 & 0.98 & 0.99 & 1.94 & 3.91 & 0.24\\
\hline
OLS & 0.09 & 0.00 & 0.35 & 0.15 & 0.23 & 0.09\\
\hline
\end{tabular}
\end{table}

%% file: figs/data-description.tex
\begin{table}

\caption{Description of the conditions we used in our data application \label{tab: data-description}}
\centering
\begin{tabular}{|c|c|c|c|}
\hline
\multicolumn{1}{|l|}{} & Additional reagent                   & Target           & Sample size   \\ \hline
Observational          & -                & -      & 853\\ \hline
Condition 1            & AKT-inhibitor  &  AKT      & 911   \\ \hline
Condition 2            & G0076          &  PKC      &723   \\ \hline
Condition 3            & Psitectorignin &  PIP2     & 810   \\ \hline
Condition 4            & U0126          &  MEK       & 799  \\ \hline
\end{tabular}
\end{table}

%% file: figs/result-pip2.tex
\begin{table}

\caption{Estimation results using PIP2 as the input variable. The left shows the results when Plcg is the response, while the right being PIP3 as response. \label{tab: result-pip2}}
\centering
\resizebox{\columnwidth}{!}{%
\begin{tabular}{|c|ccc|ccc|}
\hline
\multirow{2}{*}{} & \multicolumn{3}{c|}{PIP2 -\textgreater Plcg}                                                  & \multicolumn{3}{c|}{PIP2 -\textgreater PIP3}                                                   \\ \cline{2-7} 
                  & \multicolumn{1}{c|}{CD(GCD)}       & \multicolumn{1}{c|}{IV}               & Hybrid           & \multicolumn{1}{c|}{CD(GCD)}       & \multicolumn{1}{c|}{IV}               & Hybrid            \\ \hline
Coefficient       & \multicolumn{1}{c|}{1.88}          & \multicolumn{1}{c|}{0.42}             & 0.43             & \multicolumn{1}{c|}{-1.44}         & \multicolumn{1}{c|}{0.22}             & 0.22              \\ \hline
P value           & \multicolumn{1}{c|}{1}             & \multicolumn{1}{c|}{\textless 0.0001} & \textless 0.0001 & \multicolumn{1}{c|}{1}             & \multicolumn{1}{c|}{\textless 0.0001} & \textless{}0.0001 \\ \hline
95\% CI           & \multicolumn{1}{c|}{(-5.46, 9.21)} & \multicolumn{1}{c|}{(0.40, 0.45)}     & (0.40, 0.45)     & \multicolumn{1}{c|}{(-8.50, 5.62)} & \multicolumn{1}{c|}{(0.20, 0.25)}     & (0.19, 0.25)      \\ \hline
\end{tabular}%
}
\end{table}

%% file: figs/exp1-res.tex
\begin{table}

\caption{Estimation result for Mek$\to$ Raf. \label{tab: mek-raf}}
\centering
\begin{tabular}{|c|c|c|c|}
\hline
            & CD(GCD)          & IV               & Hybrid           \\ \hline
Coefficient & 0.94             & 0.60             & 0.63             \\ \hline
P value     & \textless 0.0001 & \textless 0.0001 & \textless 0.0001 \\ \hline
95\% CI     & (0.87, 1.00)     & (0.58, 0.62)     & (0.62, 0.65)     \\ \hline
\end{tabular}
\end{table}

%% file: ms.bbl
\begin{thebibliography}{28}
\providecommand{\natexlab}[1]{#1}
\providecommand{\url}[1]{\texttt{#1}}
\expandafter\ifx\csname urlstyle\endcsname\relax
  \providecommand{\doi}[1]{doi: #1}\else
  \providecommand{\doi}{doi: \begingroup \urlstyle{rm}\Url}\fi

\bibitem[Angrist et~al.(1996)Angrist, Imbens, and
  Rubin]{angrist1996identification}
J.~D. Angrist, G.~W. Imbens, and D.~B. Rubin.
\newblock Identification of causal effects using instrumental variables.
\newblock \emph{Journal of the American statistical Association}, 91\penalty0
  (434):\penalty0 444--455, 1996.

\bibitem[Belloni et~al.(2012)Belloni, Chen, Chernozhukov, and
  Hansen]{belloni2012sparse}
A.~Belloni, D.~Chen, V.~Chernozhukov, and C.~Hansen.
\newblock Sparse models and methods for optimal instruments with an application
  to eminent domain.
\newblock \emph{Econometrica}, 80\penalty0 (6):\penalty0 2369--2429, 2012.

\bibitem[Belloni et~al.(2018)Belloni, Chernozhukov, Chetverikov, Hansen, and
  Kato]{belloni2018high}
A.~Belloni, V.~Chernozhukov, D.~Chetverikov, C.~Hansen, and K.~Kato.
\newblock High-dimensional econometrics and regularized gmm.
\newblock \emph{arXiv preprint arXiv:1806.01888}, 2018.

\bibitem[Cameron and Trivedi(2005)]{cameron2005microeconometrics}
A.~C. Cameron and P.~K. Trivedi.
\newblock \emph{Microeconometrics: methods and applications}.
\newblock Cambridge university press, 2005.

\bibitem[Candes et~al.(2007)Candes, Tao, et~al.]{candes2007dantzig}
E.~Candes, T.~Tao, et~al.
\newblock The dantzig selector: Statistical estimation when p is much larger
  than n.
\newblock \emph{Annals of statistics}, 35\penalty0 (6):\penalty0 2313--2351,
  2007.

\bibitem[Chen et~al.(2018)Chen, Ren, Zhang, and Zhang]{chen2018two}
C.~Chen, M.~Ren, M.~Zhang, and D.~Zhang.
\newblock A two-stage penalized least squares method for constructing large
  systems of structural equations.
\newblock \emph{The Journal of Machine Learning Research}, 19\penalty0
  (1):\penalty0 40--73, 2018.

\bibitem[Dunker(2021)]{dunker2021adaptive}
F.~Dunker.
\newblock Adaptive estimation for some nonparametric instrumental variable
  models with full independence.
\newblock \emph{Electronic Journal of Statistics}, 15\penalty0 (2):\penalty0
  6151--6190, 2021.

\bibitem[Gimenez and Rothenh{\"a}usler(2021)]{gimenez2021causal}
J.~R. Gimenez and D.~Rothenh{\"a}usler.
\newblock Causal aggregation: estimation and inference of causal effects by
  constraint-based data fusion.
\newblock \emph{arXiv preprint arXiv:2106.03024}, 2021.

\bibitem[Gold et~al.(2020)Gold, Lederer, and Tao]{gold2020inference}
D.~Gold, J.~Lederer, and J.~Tao.
\newblock Inference for high-dimensional instrumental variables regression.
\newblock \emph{Journal of Econometrics}, 217\penalty0 (1):\penalty0 79--111,
  2020.

\bibitem[Hansen(1982)]{hansen1982large}
L.~P. Hansen.
\newblock Large sample properties of generalized method of moments estimators.
\newblock \emph{Econometrica: Journal of the econometric society}, pages
  1029--1054, 1982.

\bibitem[Heinze-Deml et~al.(2018)Heinze-Deml, Peters, and
  Meinshausen]{heinze2018invariant}
C.~Heinze-Deml, J.~Peters, and N.~Meinshausen.
\newblock Invariant causal prediction for nonlinear models.
\newblock \emph{Journal of Causal Inference}, 6\penalty0 (2), 2018.

\bibitem[Henderson and Searle(1981)]{henderson1981vec}
H.~V. Henderson and S.~R. Searle.
\newblock The vec-permutation matrix, the vec operator and kronecker products:
  A review.
\newblock \emph{Linear and multilinear algebra}, 9\penalty0 (4):\penalty0
  271--288, 1981.

\bibitem[Lewbel(2012)]{lewbel2012using}
A.~Lewbel.
\newblock Using heteroscedasticity to identify and estimate mismeasured and
  endogenous regressor models.
\newblock \emph{Journal of Business \& Economic Statistics}, 30\penalty0
  (1):\penalty0 67--80, 2012.

\bibitem[Lin et~al.(2015)Lin, Feng, and Li]{lin2015regularization}
W.~Lin, R.~Feng, and H.~Li.
\newblock Regularization methods for high-dimensional instrumental variables
  regression with an application to genetical genomics.
\newblock \emph{Journal of the American Statistical Association}, 110\penalty0
  (509):\penalty0 270--288, 2015.

\bibitem[M{\'a}ty{\'a}s et~al.(1999)M{\'a}ty{\'a}s, Gourieroux, Phillips,
  et~al.]{matyas1999generalized}
L.~M{\'a}ty{\'a}s, C.~Gourieroux, P.~C. Phillips, et~al.
\newblock \emph{Generalized method of moments estimation}, volume~5.
\newblock Cambridge University Press, 1999.

\bibitem[Meinshausen(2018)]{meinshausen2018causality}
N.~Meinshausen.
\newblock Causality from a distributional robustness point of view.
\newblock In \emph{2018 IEEE Data Science Workshop (DSW)}, pages 6--10. IEEE,
  2018.

\bibitem[Meinshausen et~al.(2016)Meinshausen, Hauser, Mooij, Peters, Versteeg,
  and B{\"u}hlmann]{meinshausen2016methods}
N.~Meinshausen, A.~Hauser, J.~M. Mooij, J.~Peters, P.~Versteeg, and
  P.~B{\"u}hlmann.
\newblock Methods for causal infåerence from gene perturbation experiments and
  validation.
\newblock \emph{Proceedings of the National Academy of Sciences}, 113\penalty0
  (27):\penalty0 7361--7368, 2016.

\bibitem[Newey and McFadden(1994)]{newey1994large}
W.~K. Newey and D.~McFadden.
\newblock Large sample estimation and hypothesis testing.
\newblock \emph{Handbook of econometrics}, 4:\penalty0 2111--2245, 1994.

\bibitem[Pearl et~al.(2009)]{pearl2009causal}
J.~Pearl et~al.
\newblock Causal inference in statistics: An overview.
\newblock \emph{Statistics surveys}, 3:\penalty0 96--146, 2009.

\bibitem[Peters et~al.(2016)Peters, B{\"u}hlmann, and
  Meinshausen]{peters2016causal}
J.~Peters, P.~B{\"u}hlmann, and N.~Meinshausen.
\newblock Causal inference by using invariant prediction: identification and
  confidence intervals.
\newblock \emph{Journal of the Royal Statistical Society. Series B (Statistical
  Methodology)}, pages 947--1012, 2016.

\bibitem[Pfister et~al.(2019)Pfister, B{\"u}hlmann, and
  Peters]{pfister2019invariant}
N.~Pfister, P.~B{\"u}hlmann, and J.~Peters.
\newblock Invariant causal prediction for sequential data.
\newblock \emph{Journal of the American Statistical Association}, 114\penalty0
  (527):\penalty0 1264--1276, 2019.

\bibitem[Poirier(2017)]{poirier2017efficient}
A.~Poirier.
\newblock Efficient estimation in models with independence restrictions.
\newblock \emph{Journal of Econometrics}, 196\penalty0 (1):\penalty0 1--22,
  2017.

\bibitem[Ray and Saumyadipta(2012)]{ray2012transformation}
S.~Ray and P.~Saumyadipta.
\newblock A computational framework to emulate the human perspective in flow
  cytometric data analysis.
\newblock \emph{PLoS ONE}, 7(5)\penalty0 (35693), 2012.

\bibitem[Rothenh{\"a}usler et~al.(2019)Rothenh{\"a}usler, B{\"u}hlmann,
  Meinshausen, et~al.]{rothenhausler2019}
D.~Rothenh{\"a}usler, P.~B{\"u}hlmann, N.~Meinshausen, et~al.
\newblock Causal dantzig: fast inference in linear structural equation models
  with hidden variables under additive interventions.
\newblock \emph{The Annals of Statistics}, 47\penalty0 (3):\penalty0
  1688--1722, 2019.

\bibitem[Rothenh{\"a}usler et~al.(2021)Rothenh{\"a}usler, Meinshausen,
  B{\"u}hlmann, and Peters]{rothenhausler2021anchor}
D.~Rothenh{\"a}usler, N.~Meinshausen, P.~B{\"u}hlmann, and J.~Peters.
\newblock Anchor regression: Heterogeneous data meet causality.
\newblock \emph{Journal of the Royal Statistical Society Series B: Statistical
  Methodology}, 83\penalty0 (2):\penalty0 215--246, 2021.

\bibitem[Sachs et~al.(2005)Sachs, Perez, Pe'er, Lauffenburger, and
  Nolan]{sachs2005causal}
K.~Sachs, O.~Perez, D.~Pe'er, D.~A. Lauffenburger, and G.~P. Nolan.
\newblock Causal protein-signaling networks derived from multiparameter
  single-cell data.
\newblock \emph{Science}, 308\penalty0 (5721):\penalty0 523--529, 2005.

\bibitem[Saengkyongam et~al.(2022)Saengkyongam, Henckel, Pfister, and
  Peters]{saengkyongam2022exploiting}
S.~Saengkyongam, L.~Henckel, N.~Pfister, and J.~Peters.
\newblock Exploiting independent instruments: Identification and distribution
  generalization.
\newblock In \emph{International Conference on Machine Learning}, pages
  18935--18958. PMLR, 2022.

\bibitem[Wright(1928)]{wright1928tariff}
P.~G. Wright.
\newblock \emph{Tariff on animal and vegetable oils}.
\newblock Macmillan Company, New York, 1928.

\end{thebibliography}
